\documentclass[12pt,a4paper]{article}

\usepackage[left=2cm, right=2cm, top=2cm, bottom=2cm]{geometry}
\usepackage{float}
\usepackage{titlesec}

\usepackage{url}

\usepackage{environ}
\usepackage{ifthen}

\usepackage[LGR,T1]{fontenc} 
\usepackage[greek,english]{babel}
\usepackage{scrextend}

\usepackage{calc,amsmath,amssymb,amsfonts,stmaryrd}
\usepackage{algorithmic}
\usepackage{mathtools}

\usepackage{graphicx}
\usepackage{textcomp}
\usepackage{tikz}

\usepackage{fancyhdr} 
\usepackage{colortbl} 
\usepackage{lineno} 
\usepackage{hyperref} 
\usepackage{tabularx} 
\usepackage{tcolorbox} 

\usepackage{cleveref} 
\usepackage[user]{zref}

\usepackage{eiffel}
\usepackage{lipsum}
\usepackage{fancyvrb}

\definecolor{bgcolor}{HTML}{FFFFB3}
\definecolor{bordercolor}{HTML}{777777}
\definecolor{commentcolor}{HTML}{333333}
\definecolor{headercolor}{HTML}{c90000}
\newlength{\borderwidth}
\titleformat{\section}
{\normalfont\Large\color{headercolor}}{\bf{\thesection}}{1em}{}

\renewcommand{\bf}[1]{{\textnormal{\fontfamily{qtm}\selectfont\bfseries#1}}}
\newcommand{\defnfmt}[1]{\textcolor{green!50!black}{/\textit{#1}/}}

\newcommand{\paralllel}{\textnormal{|||}}

\makeatletter
\zref@newprop{name}{}
\zref@newprop{anchor}{}
\zref@addprop{main}{name}

\newcommand{\definitionlabel}[2]{%
  \zref@setcurrent{name}{#2}%
  \zlabel{#1}%
  \hypertarget{#1}{}
}

\newcommand{\referencename}[1]{%
  \hyperlink{#1}{\defnfmt{\zref@extract{#1}{name}{}}}%
}

\newcommand{\defnraww}[2]{\definitionlabel{#1}{#2}{\defnfmt{#2}}}
\newcommand{\defnraw}[2]{{\scriptsize \defnraww{#1}{#2}}}

\newcommand{\pdomain}[1]{\underline{\bl{#1}}}
\newcommand{\prange}[1]{\textoverline{\bl{#1}}}

\newcommand{\defn}[2]{\raggedleft\defnraw{#1}{#2}}

\newcommand{\bff}[1]{\textnormal{\bf{#1}}}

\newcommand{\bll}[1]{\textcolor{blue}{#1}} 
\newcommand{\bl}[1]{\bll{\textit{#1}}} 

\newcommand{\textoverline}[1]{$\overline{\mbox{#1}}$} 

\newcommand{\ifelse}[3]{\bl{\bff{if}\textit{ #1 }\bff{then}\textit{ #2 }\bff{else}\textit{ #3 }\bff{end}}} 

\newcommand{\aloop}[1]{\bl{\bff{loop}\textit{ #1 }\bff{end}}} 
\newcommand{\fromloop}[3]{\bl{\bff{from}\textit{ #1 }\bff{until}\textit{ #2 }\bff{loop}\textit{ #3 }\bff{end}}} 
\newcommand{\whileloop}[2]{\bl{\bff{while }\textit{#1 }\bff{loop}\textit{ #2 }\bff{end}}} 
\newcommand{\repeatloop}[2]{\bl{\bff{repeat }\textit{#2 }\bff{until}\textit{ #1 }\bff{end}}} 

\newcommand{\pr}[1][\relax]{\ifx\relax#1\bl{p} \else\bl{p$_{#1}$}\fi} 
\newcommand{\St}[1][\relax]{\ifx\relax#1\bl{S} \else\bl{S$_{#1}$}\fi} 
\newcommand{\Pre}[1][\relax]{\ifx\relax#1\bl{Pre} \else\bl{Pre$_{#1}$}\fi} 
\newcommand{\post}[1][\relax]{\ifx\relax#1\bl{post} \else\bl{post$_{#1}$}\fi} 

\newcommand{\pwrong}{False}
\newcommand{\pright}{True}

\newcommand{\ToDo}[1][\relax]{\ifx\relax#1\textcolor{red}{\bf{ToDo.}} \else\textcolor{red}{\bf{ToDo: #1}}\fi}

\newcommand{\refines}{\sqsubseteq} 
\newcommand{\specializes}{\subseteq} 
\makeatletter
\newcommand{\arrowbar}{}
\newcommand{\pgets}{}

\DeclareRobustCommand{\arrowbar}{\mathrel{\mathpalette\p@to@gets\to}}
\DeclareRobustCommand{\pgets}{\mathrel{\mathpalette\p@to@gets\gets}}

\newcommand{\p@to@gets}[2]{%
  \ooalign{\hidewidth$\m@th#1\mapstochar\mkern5mu$\hidewidth\cr$\m@th#1\to$\cr}%
}
\makeatother

\tcbset{
	defbox/.style={
		arc=0pt,
		before=\vspace{-1pt},
		after=\relax,
		colback=bgcolor,
		colframe=bordercolor,
		toprule=0pt,
		rightrule=\borderwidth,
		bottomrule=0pt,
		leftrule=\borderwidth,
		left=10pt,
		right=10pt,
		top=5pt,
		bottom=5pt,
		boxsep=0pt,
		width=\textwidth
	},
}

\newenvironment{bcolfr}[1][\relax]{
	\noindent
	\ifstrequal{#1}{\relax}{\tcolorbox[defbox]}{
		\ifstrequal{#1}{bot}{\tcolorbox[defbox, bottomrule=\borderwidth]}{
			\tcolorbox[defbox, toprule=\borderwidth]}}
	\ifstrequal{#1}{bot}{\vspace{0.5em}}{{\centering\bf{#1}\par\vspace{0.5em}}}
}{
	\endtcolorbox%
}

\newenvironment{colfr}[1][\relax]{
	\vspace{0.5em}\noindent
	\begin{tcolorbox}[
		arc=0pt,
		colback=bgcolor,
		colframe=bordercolor,
		boxrule=\borderwidth,
		left=10pt,
		right=10pt,
		top=5pt,
		bottom=5pt,
		boxsep=0pt,
		width=\textwidth
		]
	\ifx\relax#1 \relax\else {\centering\bf{#1}\\\vspace{0.5em}}\fi
	}{%
	\end{tcolorbox}%
	\vspace{0.5em}%
}

\newenvironment{sdef2}[4]{
	\noindent
	\begin{minipage}[c]{(#1\textwidth)}%
		\def\definitionend{%
		\end{minipage}%
		\hfill
		\begin{minipage}[c]{\dimexpr(#2\textwidth)-\columnsep\relax}%
			\defn{#3}{#4}%
		\end{minipage}%
	}%
}{%
	\definitionend%
}

\newenvironment{sdef}[3]{
	\noindent
	\begin{minipage}[c]{(#1\textwidth)}%
		\def\definitionend{%
		\end{minipage}%
		\hfill
		\begin{minipage}[c]{\dimexpr\textwidth-(#1\textwidth)-\columnsep\relax}%
			\defn{#2}{#3}%
		\end{minipage}%
	}%
}{%
	\definitionend%
}

\newenvironment{definition}[3][0.66]{
	\begin{colfr}
		\begin{sdef}{#1}{#2}{#3}
	}{%
		\end{sdef}
	\end{colfr}
}

\newenvironment{note}[1]{
	\begin{addmargin}[0em]{0em}
	\textit{#1:}
}{
	\end{addmargin}
	\vspace{1ex}
}

\title{Meanings as programs:\\ Programming Really Is Simple Mathematics}
\author{Bertrand Meyer, Reto Weber}

\begin{document}

\vspace*{0.2cm}

\begin{center}
	{\LARGE \bfseries Meanings as programs:}\\[0.5em]
	{\LARGE \bfseries Programming Really Is Simple Mathematics}\\[1.5em]
	{\Large Bertrand Meyer, Reto Weber}\\
    \vspace{0.2cm}{\normalsize Constructor Institute of Technology}

\end{center}
\vspace{0.4cm}
\setlength{\leftskip}{5.8cm}
\noindent\footnotesize{\textit{Dedicated to Yuri Gurevich on the occasion of his 85th birthday, in homage to his seminal contributions to the theory and practice of software.}}

\vspace{0.2cm}

\begin{abstract}
	\noindent A re-construction of the fundamentals  of programming as a small mathematical theory (``PRISM'') based on elementary set theory. Highlights:

	\begin{itemize}
		\item Zero axioms. No properties are assumed, all are proved (from standard set theory).
		\item A single concept covers specifications and programs.
		\item Its definition only involves one relation and one set.
		\item Everything proceeds from three operations: choice, composition and restriction.
		\item These techniques suffice to derive the axioms of classic papers on the “laws of programming” as consequences and prove them mechanically.
		\item The ordinary subset operator suffices to define both the notion of program correctness and the concepts of specialization and refinement.
		\item From this basis, the theory deduces dozens of theorems characterizing important properties of programs and programming.
		\item All these theorems have been mechanically verified (using Isabelle/HOL); the proofs are available in a public repository.
	\end{itemize}

	\noindent This paper is a considerable  extension and rewrite of an earlier contribution \cite{meyer2013}.

\end{abstract}

\section{Overview} \label{overview}

\setlength{\leftskip}{0pt}

The present article is a development of a long-term project to establish computer programming as a mathematical theory. The core ideas were outlined in reference \cite{meyer2013}, of which it is a revision and extension, with all properties now formally verified.
\begin{note}{Caveat}
	The construction of the theory, starting with section \ref{basic}, assumes no prior knowledge of programming. While readers may have such knowledge (having possibly, for example, encountered the term ``bug'' before discovering its definition in section \ref{feasibility}), the best way to approach this exposition is most of the time to pretend to be  reading about a new subject --- occasionally breaking this rule by relating the formal view (with the help of ``Explanation'' and ``Justification'' paragraphs) to one's existing understanding of programming concepts.
\end{note}


\subsection{Programming and mathematics: two worlds, or one?} \label{twoworlds}

Anyone who has both written programs and studied elementary mathematics will have noticed how close the business of writing programs is to the business of proving theorems.

\begin{figure}[hbt!]
	\centering
	\includegraphics[width=0.60\linewidth]{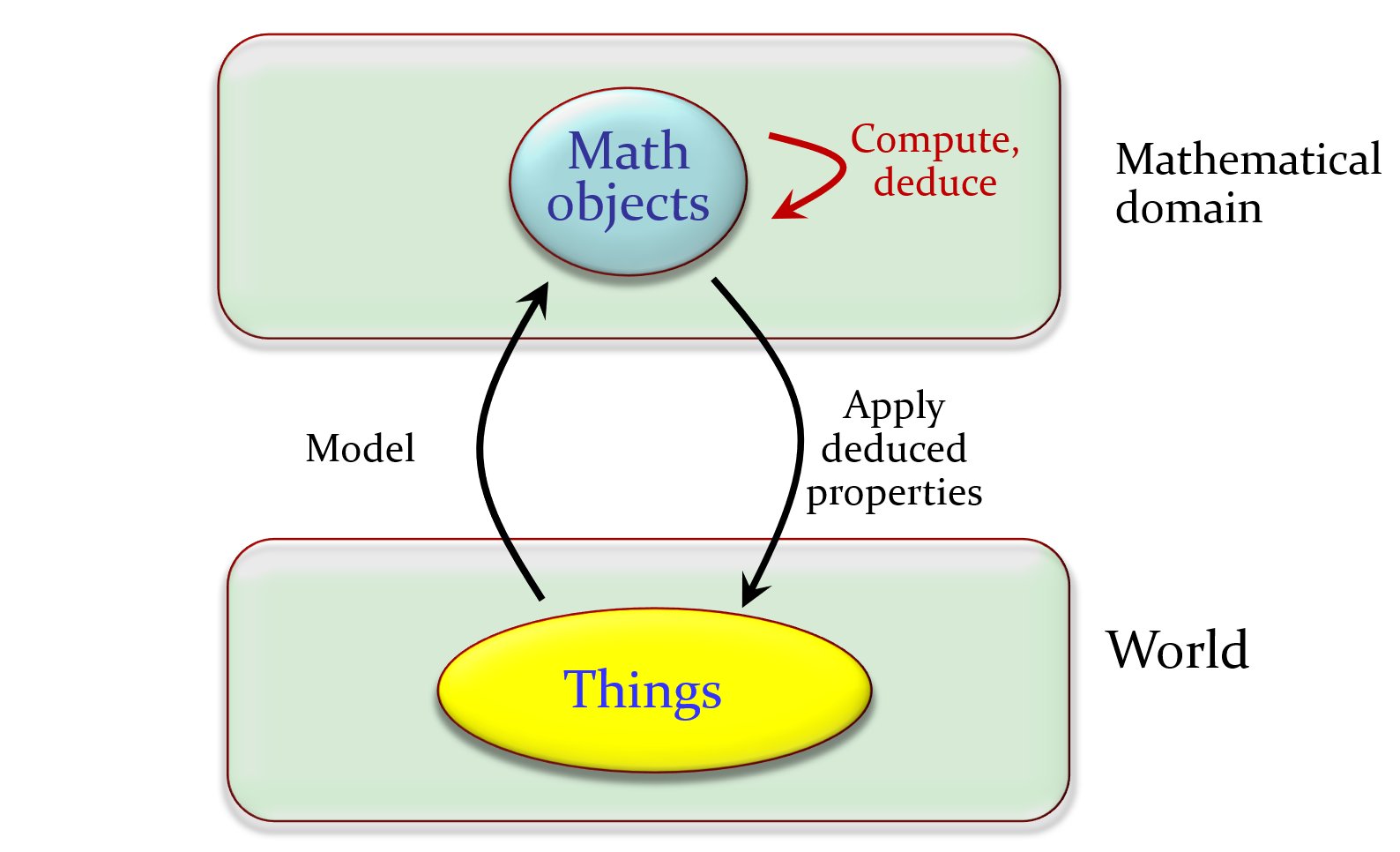}
	\caption{\bf{Modeling through mathematics}}\label{through}
\end{figure}

Instead of taking advantage of this closeness, however, most of the work relating programs to mathematics proceeds as if the two worlds were separate. It typically considers a program as if it were some pre-existing object of study, like a bird for an ornithologist or a wave for a physicist, and sets out to explain it in terms of mathematical concepts. Such an approach follows the time-honored process of using  mathematics in the natural sciences: devise a mathematical model of some natural  objects;  work on the model to deduce interesting properties; then transpose these properties back to the original objects of interest (Fig. \ref{through}). Floyd's foundational semantics paper \cite{floyd1967} was accordingly titled ``\textit{Assigning Meanings to Programs}''.

The B book \cite{abrial1996} provocatively and famously reversed the pairing, promising to ``\textit{Assign Programs to Meanings}'': write very abstract programs with clear mathematical properties, then refine them progressively into more concrete programs until they are deemed practical enough for direct use.

We go one step further towards simplification by \textit{removing any distinction} between program and meaning. The corresponding slogan could be ``Meanings As Programs''. A program is  a mathematical object, from a (simple) theory, which happens to be suitable  both for execution by a computer and for perusal, under a suitably designed notation (``syntax''), by human readers.


\subsection{Staying in the world of mathematics} \label{staying}

\begin{figure}[hbt!]
	\centering
	\includegraphics[width=0.60 \linewidth]{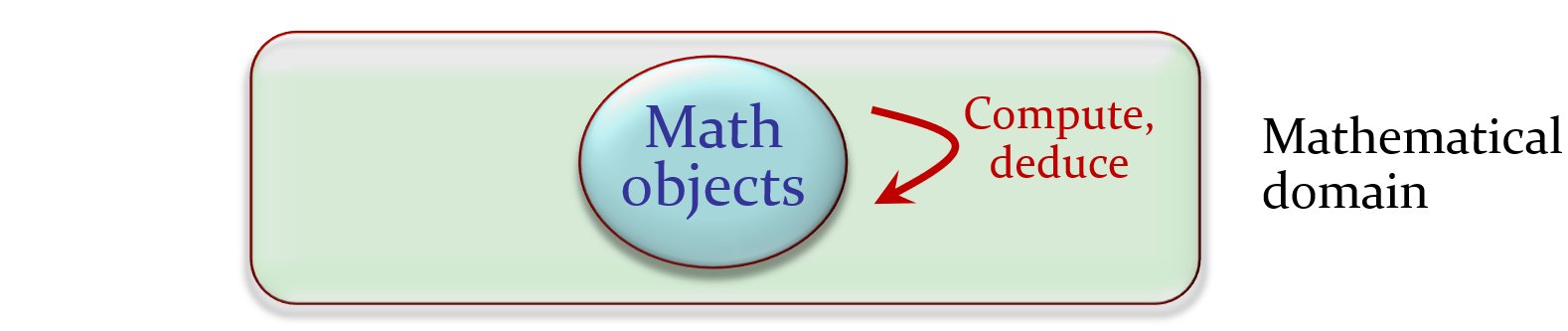}
	\caption{\bf{Reasoning within mathematics}}\label{math}
\end{figure}

The key word in the last sentence of the previous paragraph is ``is'': the mathematical object does not provide a model of some thing endowed with its own existence, it \textit{is} that thing. Unlike the two-world model of Fig. \ref{through}, applicable to the natural sciences, the process resembles the simpler one-world model of pure mathematics, which only includes (Fig. \ref{math}) the top part of the earlier schema.

\noindent In other words, we start in the world of mathematics, not programming, and do not ever leave it, even when we go down to concepts that typically appeal to programmers rather than pure mathematicians.

As one does in mathematics, we construct a theory in the usual way, by defining some mathematical objects and proving that they possess certain properties. That theory happens to cover the concepts of programming and its objects are programs, data structures, instructions and other artifacts of programming. As they come from a mathematical theory,  they satisfy well-defined properties, proved as theorems.

One difference with the usual (non-software-related) practice of mathematical work has to do with notation. While mathematicians pay considerable attention to notations, they do not generally have the  programmer's needs for very large formal texts. In a mathematical article or book, the formal parts  each spread over a few lines or, rarely, a few pages, much less than the thousands to millions of lines taken up by a program text, from which concerns arise that are unknown to mathematicians: we need modular constructs (routines, classes, packages...) to organize programs into manageable pieces; to resolve possible ambiguity, we cannot rely on the common sense of a human reader (who readily sees, for example, that one formula extends over two lines but the third line starts a new formula), and must rely instead on rigorously defined syntax, with specific properties (context-free, LL(1), LALR...) enabling compilers to process the texts. To make programs at least moderately readable, debuggable and maintainable, the programming language --- even if it is of the esoteric, geeky kind, for example C or Python --- must retain some of the verbosity of natural language (usually English) and even some of its words for use as keywords.

As the present work will show, however, programming languages are in the end just notations (``syntactic sugar'' is the accepted phrase) for mathematical objects. For example, a standard mathematical concept is the \textit{restriction} of a relation: \bl{\bl{C$:$ }r}, for a set \bl{C} and a relation \bl{r}, is the relation identical to \bl{r} except that its domain has been restricted to \bl{C}. For a programming audience, we allow writing it as \bl{\bf{if} C \bf{then} r \bf{end}} (see section \ref{conditionals}). That form is only an alternative notation for the classical mathematical concept of restriction.

Using this general idea that programs and their components are objects of a mathematical theory couched in a programmer-ready notation, we can pursue the goal of constructing a fully developed programming language, or \textit{re-}constructing an existing language (at least one whose semantics is sufficiently simple and regular), in the form of programming-oriented representations of objects of the mathematical theory.

The present article falls short of attaining this goal, but moves in its direction, up to the fundamental control structures of modern programming languages (sequence, various forms of conditional instruction, various forms of repetition and loops), basic Design by Contract concepts (preconditions, postconditions), and concurrency. The exposition proceeds (starting with section \ref{basic}) according to the principles defined above: build a theory, starting from extremely simple notions (a relation and a function), derive and prove (with mechanical verification through Isabelle/HOL) theorems capturing important properties of  its mathematical objects, and provide programming-like syntax --- specifically, Eiffel-like syntax, since at the end of the journey we might reach the entire language --- to express them.


\subsection{How not to  use mathematics for programming (1)} \label{kahn}

The present work is in part a reaction against some traditional approaches which go in the programs-to-math direction rather than math-to-programs as developed here. Two well-known contributions by prestigious authors typify what we are up against.

Consider first Gilles Kahn's classic ``Natural Semantics'' paper \cite{kahn1987}. It specifies a simple programming language and has a list of definitions for programming constructs, assuming a considerable mathematical baggage. The semantics of the conditional expression (``if ... then ... else ...''), for example, is specified through two inference rules:


\begin{colfr}
    	$$\frac{\rho \vdash E_1 \Rightarrow True \hspace{3em} \rho \vdash E_2 \Rightarrow \alpha}{\rho \vdash \text{\bf{if} } E_1 \text{ \bf{then} } E_2 \text{ \bf{else} } E_3 \Rightarrow \alpha}$$
        \vspace{1em}
    	$$\frac{\rho \vdash E_1 \Rightarrow False \hspace{3em} \rho \vdash E_3 \Rightarrow \alpha}{\rho \vdash \text{\bf{if} } E_1 \text{ \bf{then} } E_2 \text{ \bf{else} } E_3 \Rightarrow \alpha}$$
        \centering
    	
\end{colfr}
\vspace{-2em}
{\begin{center}
    \bf{Conditional expression: ``natural'' (?) semantics, from \cite{kahn1987}}\label{fig:natural}
\end{center}}

\noindent (Explanation of this rule: its purpose is to specify how to evaluate, during program execution, the conditional expression \bl{\bf{if} E1 \bf{then} E2 \bf{else} \bl{E3}}. We characterize the state of the program, at the time of evaluation, by a  ``variable binding'' \bl{$\rho$}, which maps names of variables to their values at that time; for example, after an assignment of the value True to the variable \bl{x}, the binding $\rho$ must be such that, under $\rho$, \bl{x} is bound to True. Such a property is written \bl{$\rho$ $\vdash$ x $\Rightarrow$ True}, read aloud as ``\textit{under} binding \bl{$\rho$}, the variable \bl{x} \textit{evaluates to} \bl{True}''. The rule of  Fig. \ref{fig:natural} has two parts, each of which is an ``inference rule'' telling us that if the properties above the horizontal line (the hypotheses) hold we may deduce the property below the line (the conclusion). This particular inference rule states that if, under the binding \bl{$\rho$},  \bl{E1} evaluates to True, and under that same binding \bl{E2} evaluates to some value \bl{$\alpha$}, then, again in that binding \bl{$\rho$}, the whole expression \bl{\bf{if} E1 \bf{then} E2 \bf{else} E3} will also evaluate to \bl{$\alpha$}. And the second rule gives us the corresponding property for the False case. Wow!)

\begin{figure}[hbt!]
	\centering
	\includegraphics[width=0.25 \linewidth]{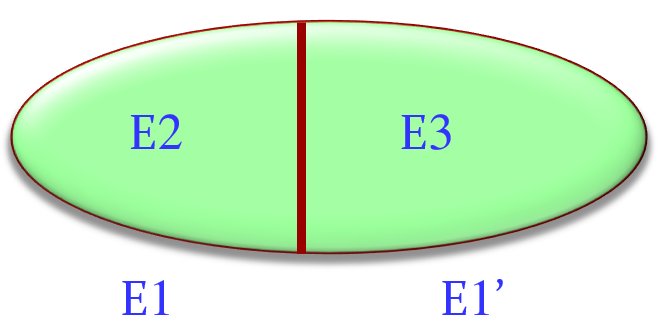}
	\caption{\bf{Visualizing the evaluation of a conditional expression or instruction}}\label{fig:conditional}
\end{figure}

\begin{figure}[hbt!]
	\centering
	\includegraphics[width=0.12 \linewidth]{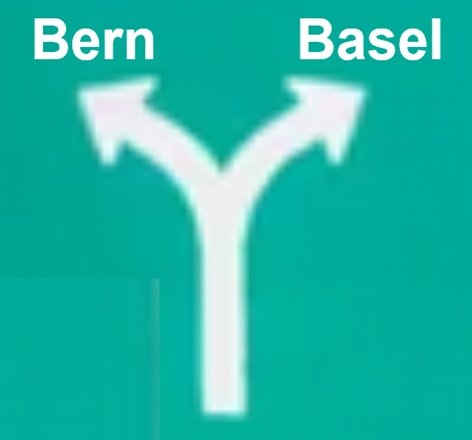}
	\caption{\bf{A "conditional" road sign}}\label{fig:fork}
\end{figure}

One can only shake one’s head: the notion of conditional expression can be explained in full to a 6-year-old: wherever \bl{E1} holds use \bl{E2}, otherwise use \bl{E3}. You may summon Fig. \ref{fig:conditional} for help. Or realize how easy it is to interpret an ``if... then... else'' road sign (Fig. \ref{fig:fork}) even though most drivers would be surprised to hear that if they want to avoid ending up in Bern when they would actually like to get to Basel they first have to study Boolean algebra, variable bindings and inference rules.

If modeling such a basic construct --- in a toy functional programming language, which in the cited paper does not even have assignments, routines or loops --- requires such a heavy conglomerate of concepts and notations, what would we need for the truly hard mechanisms of programming?  That cannot be right. The definition of conditional obtained below will conform to what the last figure suggests: take the \textit{union} of \bl{E2} and \bl{E3}, each \textit{restricted} respectively to \bl{E1} and its complement.

\subsection{How not to  use mathematics for programming (2)} \label{hoare}

The second example of how not to approach the relationship between programming and mathematics is another milestone paper, this one by Tony Hoare and eight co-authors \cite{hoare1987}, extended and refined over the next 25 years later with other coauthors \cite{hoare2012}. It introduces ``Laws of Programming'' intended to serve as a basis for the entire discipline.

There is much to admire in this effort, but one can only balk at the general methodology: \textit{postulating} desired laws as axioms. Some of these laws, together with their comments, appear in Fig. \ref{choicelaws}; they are the ones dealing with the cited authors' ``choice'' operator, for which they use the union symbol $\cup$ (or possibly some sign looking very much like it).


\begin{colfr}
\centering
    \begin{minipage}{0.75\textwidth}
	\centering
        \footnotesize
        (1) Clearly, it does not make any difference in what order a choice is offered. ``Tea or coffe?'' is the same as `` coffee or tea?''
        $$P \cup Q = Q \cup P \text{\hspace{1cm}(symmetry)}$$
        (2) A choice between three alternatives (tea, coffee, or cocoa) can be offered as first a choice between on alternative and the other two, followed (if necessary) by a choice between the other two, and it does not matter in which way the choices are grouped.
        $$P \cup (Q \cup R) = (P \cup Q) \cup R \text{\hspace{1cm}(associativity)}$$
        (3) A choice between one thing and itself offers no choice at all (Hobson`s choice).
        $$P \cup P = P \text{\hspace{1cm}(idempotence)}$$
        (4) The ABORT command already allows completely arbitrary behavior, so an offer of further choice makes no difference to it.
        $$\bot \cup P  = \bot \text{\hspace{1cm}(zero }\bot\text{)}$$
    \end{minipage}
\end{colfr}
\vspace{-2em}
{\begin{center}
    \bf{``Laws of choice'', from \cite{hoare1987}}\label{choicelaws}
\end{center}}

\noindent If we cease for an instant to believe the authors, who tell us that \bl{$\cup$} is some newly invented operator on programs, called ``choice'', but chose instead to believe our own lying eyes, which tell us that \bl{$\cup$} is  \bl{$\cup$} , meaning union, then the ``laws'' enunciated here are trivial; they are just the properties of union, which we learned in kindergarten. (We have to assume that  ``\bl{$\bot$}'' in this context is but a fancy name for the universal set.)


That is not what the article says. It is taking an entirely axiomatic approach, stating at the outset: ``\textit{we suggest that a comprehensive set of algebraic laws serves as a useful formal definition (axiomatization) of a set of related mathematical notations, and specifically of a programming language}''\cite{hoare1987}. In other words, define a new mathematical domain through a set of laws, unrelated to anything that already exists. It is inevitable, however, to ask whether an operation that is commutative like the union of subsets, associative like union, idempotent like union, has a unit like union (as well as a zero like union), and for which the authors have chosen what seems to be the same symbol, might not possibly \textit{be} union.

This line of articles includes an amazing number of such proudly shoehorned ``laws'' that turn out, if interpreted in the framework of elementary set theory, to be obvious properties. As another example, \cite{hoarepraise} has a ``law of consequence'' stating (parentheses added here for readability): \\

\bl{$($p $\refines$ \bl{p'}$)$ ~ $\land$ ~ $($$($p' $;$ q$)$ $\refines$ r$)$ ~~ $\rightarrow$ ~~ $($p $;$ q$)$ $\refines$ r} \\

\noindent where ``\bl{$\refines$}'' is refinement and ``\bl{$;$}'' is sequential composition. In mathematics, however, ``\bl{$\subseteq$}'' is subset and ``\bl{r $;$ s}'' is a notation (an alternative to \bl{s $\circ$  r}) for the composition of relations \bl{r} and \bl{s} in this order. In such a framework, this ``law of consequence (pre)'' is a trivial theorem. (See a precise form, \referencename{composespecialsafety}, in section \ref{specialsafe}.)


\subsection{Other law-guided approaches} \label{otherapproaches}

Denotational semantics \cite{milne} is the quintessential approach considering programs and programming languages as natural objects deserving analysis: for every programming construct \bl{c} it defines a ``meaning'' \bl{M} $\llbracket$\bl{c$ \rrbracket$} defined recursively, starting with special ``semantic domains''.

Seminal  books by Hehner \cite{hehner}, Morgan\cite{morgan1990, morgan}  and Back \cite{back2012} have devised theories of programming and elucidated many issues. They follow the style also found in the Hoare work cited above: define programming  constructs through their properties, in particular refinement. To discuss the conditional instruction, for example, chapter 4 of \cite{morgan} presents a syntax for the construct, provides an informal operational explanation of its  semantics, then throws in a ``refinement'' law, with no argument that it is sound with respect to some model or consistent with other laws.


\subsection{A public incitation to crime?} \label{crime}
The authors of the Hoare-style, axiom-based line of work (pursued in successive articles including \cite{hoare1987}, \cite{hoare1997}, \cite{hoare1987}, \cite{hoare2012}, \cite{hoareslides}, \cite{hoarepraise}, \cite{hoare2018}) proudly advertise its benefits, with \cite{hoareslides} invoking Bertrand Russell:

\begin{itemize}
	\item []\textit{The method of postulation has many advantages.  They are the same as the advantages of theft over honest toil.}

\end{itemize}

\noindent After reading this citation twice and checking that one is awake, the only reaction --- tempered only by the out-and-out impenetrability of  British humor to foreigners --- is to wonder what kind of argument it offers. If such theft is to be commended, mathematics becomes easy: I can postulate that P is equal to NP (or just as well, if my taste goes the other way, that they are not equal) and win the \$1-million Clay prize. Mathematics is not about postulating all you need but (as Russell brilliantly demonstrated when he was doing actual work rather than  indulging in  dubious professor jokes) to \textit{minimize} the number of axioms and, from them, \textit{prove interesting theorems}, maximizing their number.

A fundamental precept of all sciences, not just mathematics, encourages us to strive for the smallest possible set of hypotheses and to reject unnecessary assumptions (such as specially introduced constants in physics). In the history of science it has been expressed as ``Occam's Razor'', Newton's and Laplace's ``\textit{we are to admit no more causes of natural things than such as are both true and sufficient to explain their appearances}'',  Einstein's well-known maxim about simplicity etc.

The number of postulated axioms in the PRISM theory presented below is, exactly, zero. We are happy to start from the established properties of set theories, whose axioms we take from credible sources (found in well-known textbooks such as \cite{mendelson}), and set out to prove the properties of programs.

Rejecting axioms, or minimizing their number, is not just a methodological guideline justified by striving for elegance. There is a downside to ``postulating'' what you want: how do you know the axioms thrown into the kitchen sink are all \textit{consistent} with each other? Consistency proofs are difficult, and when we are dealing with sophisticated constructions such as occur in computer science often teeter on the brink of undecidability. Dijkstra, who cites Hoare's citation of Russell's provocation \cite{dijkstra1974}, raises this question even though he is always deferential to Hoare:

\begin{itemize}
	\item []
	      \textit{A possibly more valid doubt may be raised by the question ``How do you know that your axioms are consistent?''.}

	      \textit{My inclination --- but am I perhaps awfully naive? --- is to say ``Well, I am pretty sure that they are consistent. And if they are not, well that is my risk! The worst that I could have done is talking about an empty universe, and, although fruitless, that cannot have done much harm either.''}
\end{itemize}

\noindent  The reader will appreciate the sudden ``not much harm'' comment from the usually stern  disciplinarian of programming methodology, not known for being indulgent towards imprecision. We should remember that a primary goal of applying formal methods in software engineering is to help ensure that future software does not reproduce the behavior of Therac-25 or Ariane 5.

The dangers of an axiomatic approach, to which Dijkstra timidly alluded, are not just theoretical. Shaoying Liu, studying Morgan's laws of refinement (\cite{morgan}), pointed out  \cite{Shaoying} that they make it possible to refine a feasible operation into an infeasible one\footnote{In the PRISM framework developed in this article, the infeasible version is, per the formal definitions of section \ref{refinedef}, a refinement but not an implementation of the original version.}. The lack of a precise mathematical specification and proofs opens the way to such consequences, and does not allow determining whether they are intentional --- and then how they one should handle them  --- or a mistake in the theory. (``Feature or bug?'')

The only way to avoid such predicaments is to move away from axiom inebriation and, having woken up sober, to shift the focus to theorems and their mechanically-checked proofs.


\subsection{A sober approach} \label{sober}

Beyond the natural glee that comes from witnessing great men getting entangled in questionable pronouncements, the lesson of the preceding analysis is twofold:
\begin{itemize}
	\item In analyzing software we should not pretend programs pre-exist their mathematical models but consider that they \textit{are} mathematical objects.

	\item A suitable theory should be ``as simple as possible but not simpler'' in the following sense: \textit{simple programming concepts should have simple mathematical descriptions}. There are complex concepts in software, and we may expect their descriptions to be non-trivial too. But a conditional expression --- yes, it is a straightforward concept and no, it cannot possibly require heaps of advanced mathematical concepts (inference rules, bindings, deductions...) to be modeled mathematically. To paraphrase Alan Kay\footnote{Kay's oft-quoted aphorism is: ``\textit{Simple things should be simple, complex things should be possible}''.}, simple things should have simple specifications, and complex things should have specifications.
\end{itemize}

\noindent The rest of this article presents the starting elements of a theory of software based on elementary (high-school-level) set-theoretical concepts --- intersection, union, subset, relation, function, composition... --- and no new axiom whatsoever. It defines programming constructs as mathematical objects utilizing these concepts, and \textit{proves} all the often remarkable properties that they possess. All the properties and their proofs have been mechanically checked; they are recorded in a repository, which anyone can use to reproduce the proofs. As far as we have been able to determine, all the ``laws'' and ``axioms'' of the cited articles by Hoare and colleagues are theorems in the present framework, including laws of concurrency such as the ``exchange axiom'' and associated exchange laws (see section \ref{concurrencyproperties}).

The PRISM approach of this article leaves the matter of consistency to mathematicians (from Zermelo and Frankel to Gödel and Cohen) and starts from their axioms, proving all new properties along the way.

Much of the formal work in software is axiom-rich and theorem-poor. One builds a complicated mathematical basis for a fairly small Return On Investment. Since that basis cannot be checked for consistency, it is fraught with uncertainty (the reverse of what we normally expect from mathematics!): over the entire edifice hangs the specter of contradiction. Instead, the present approach is not only axiom-poor but \bf{axiom-free}, and \bf{theorem-rich}, hoping to deliver a good ROI to any practicing programmer willing to learn a bit of simple theory.

\section{Notations} \label{notation}

The discussion relies on standard concepts of elementary set theory and associated notations, sometimes adapted.
\begin{itemize}
	\item []
	      (\bf{Short version of section \ref{notation}}. A reader familiar with elementary set theory and eager to move to programming concepts  may simply skip to section \ref{basic} after noting these specifics:  \bl{$\{$x$:$ A $ | ~ p$ $($x$)\}$} to define a set by comprehension, with a colon and a vertical bar, also for \bl{$\forall$} and \bl{$\exists$} predicates; \bl{\underline{r}} and \bl{\textoverline{r}} for the domain and range of a relation \bl{r}$;$ and \bl{r $;$ s} for the composition of two relations in the order of their application.)
\end{itemize}


\subsection{Logic and sets} \label{logicsets}

\noindent The boolean operators are \bl{$\land$} (and), \bl{$\lor$} (non-exclusive or), \bl{$\lnot$} (not) and \bl{$\rightarrow$} (implies, also written \bl{$\Rightarrow$}). The usual operations are available on sets: \bl{$\in$} (member), \bl{$\cap$} (intersection), \bl{$\cap$} (union), \bl{$\subseteq$} (subset or equal), \bl{$\times$} (Cartesian product). Set difference uses a minus sign, ``\bl{$-$}''\footnote{The backslash symbol ``$\backslash$'' has another meaning, corestriction of relations, as seen a few paragraphs down.}. Basic sets include \bl{$\mathbb{N}$} (natural integers), \bl{$\mathbb{Z}$} (all integers) and \bl{$\mathbb{B}$} (booleans). The cardinal (number of elements) of a finite set \bl{A} is written \bl{$|$A$|$}. \bl{$\mathbb{P}$} \bl{$($A$)$}, the powerset of a set \bl{A}, is the set of all its subsets.  Similar conventions apply to universally and existentially quantified predicates: \bl{$\forall$} \bl{x$:$ A $|$ p $($x$)$} (all members of \bl{A} satisfy \bl{p}$)$, \bl{$\exists$} \bl{x$:$ A $|$ p $($x$)$} (some member of \bl{A} satisfies \bl{p}$)$\footnote{A dot it often used instead of the vertical bar, but the latter is more visible and avoids confusion for when the theory reaches object-oriented programming.}.


\subsection{Relations and functions} \label{relations}

A relation over sets \bl{A} and \bl{B} is a subset of \bl{A} $\times$ \bl{B}: a set of pairs \bl{$\{[$$x_1$, $y_1$$]$, $[$$x_2$, $y_2$$], ...\}$} where \bl{$x_i$} $\in$ \bl{A} and \bl{$y_i$} $\in$ \bl{B}.  An alternative notation for the set \bl{$\mathbb{P}$} (\bl{A $\times$ B}$)$ of relations between \bl{A}  and \bl{B} is \bl{A} \bl{\bl{$\leftrightarrow$}} \bl{B}. The source set of a relation  in \bl{A } \bl{\bl{$\leftrightarrow$}} \bl{B} is \bl{A} and its target set is \bl{B}. The inverse of a relation \bl{$r$} is written \bl{$r^{-1}$};  if \bl{r} is in  \bl{A}  \bl{$\leftrightarrow$} \bl{B}, then  \bl{$r^{-1}$}  is in \bl{B}   \bl{$\leftrightarrow$}  \bl{A} and its definition is that  \bl{$[$x, y$]$} $\in$ \bl{$r^{-1}$} if and only if \bl{$[$y, x$]$} $\in$ \bl{r}. The domain of \bl{r}, written  \underline{\bl{r}}, is the set of elements \bl{x} of \bl{A} such that \bl{r} contains a pair \bl{$[$x, y$]$} for some \bl{y} in \bl{B}; its range, written  \bl{\textoverline{r}}, is the set of \bl{y} of \bl{B} such that \bl{r} contains a pair \bl{$[$x, y$]$} for some \bl{x} in \bl{A}\footnote{To avoid any confusion between source set and domain, and between target set and range: stating that a relation \bl{r}  is in \bl{A } \bl{\bl{$\leftrightarrow$}} \bl{B}, with domain \bl{A} and range \bl{B}, plays the role of a declaration for \bl{r}, specifying the sets from which first and second elements (respectively$)$ of pairs in the relation \textit{may} take their values. The domain and range specify which elements \textit{do} appear in \bl{r}'s actual pairs.}. Clearly, \bl{\underline{r} $\subseteq$ A}, \bl{\textoverline{r} $\subseteq$  B},  \bl{\underline{r}} = \bl{\textoverline{r$^{-1}$}} and \bl{\textoverline{\bl{r}}} = \bl{\underline{r$^{-1}$}}.

	For a relation \bl{r}  in \bl{A } \bl{\bl{$\leftrightarrow$}} \bl{B} and a subset \bl{X} of \bl{A}, the image of \bl{X} by \bl{r} is written simply \bl{r $($X$)$}; it is the subset of the target set \bl{B} made of all the \bl{y} such that \bl{r} contains a pair \bl{$[$x, y$]$} for some \bl{x} in \bl{A}. Properties include \bl{r $($$\varnothing$$) = \varnothing$}, \bl{r $($X'$)$ $\subseteq$ r $($X$)$} whenever \bl{X' $\subseteq$ X} $($referred to in later sections as \defnraww{imagesubset}{Image\_subset}$)$, \bl{r $($X$)$ $\subseteq$ \textoverline{r}} for any \bl{X} $($\defnraww{subsetimage}{Subset\_image}$)$, \bl{r $($X $\cup$ Y$)$ = r $($X$)$ $\cup$ r $($Y$)$} $($\defnraww{imageunion}{Image\_union}$)$, \bl{r $($X $\cap$ Y$)$ $\subseteq$ r $($X$)$ $\cap$ r $($Y$)$} $($\defnraww{imageinter}{Image\_inter}$)$, and \bl{r $($A$)$ = r $($\underline{r}$)$ = \textoverline{r}}. A relation \bl{r} is total if \bl{r $($X$)$} is non-empty for non-empty \bl{X}. $($Two equivalent definitions are: $($1$)$ {\underline{\bl{r}} =  \bl{S}} $;$ $($2$)$ for any \bl{x} in \bl{A}, \bl{r} contains at least one pair of the form \bl{$[$x, y$]$} for some \bl{y}.$)$

	The restriction \bl{r / X} and corestriction \bl{r $\backslash$ Y} of a relation \bl{r}  in \bl{A } \bl{\bl{$\leftrightarrow$}} \bl{B}, for subsets \bl{X} of \bl{A} and \bl{Y} of \bl{B},  are the subsets of \bl{r} containing only the pairs \bl{$[$x, y$]$} for which, respectively, \bl{x} $\in$ {X} and \bl{y} $\in$ {Y}. Clearly, \bl{\underline{r / X} $\subseteq$  X} and \bl{\textoverline{r $\backslash$ Y} $\subseteq$  Y}. More precisely, \bl{\underline{r / X}} =  \bl{\underline{r}} $\cap$ \bl{X} and \bl{\textoverline{r $\backslash$ Y}} =  \bl{\textoverline{r} $\cap$ Y} $($\defnraww{domainrestrict}{Domain\_restrict}$)$. Also, \bl{\textoverline{r $($X$)$} = \textoverline{r / X} $\subseteq$ \textoverline{r $($A$)$}} $($\defnraww{imagerestrict}{Image\_restrict}$)$.

	Composition of relations uses the operator ``\bl{$;$}'' which lists its operands in the order of application: \bl{r $;$ s} is the set of pairs \bl{$[$x, z$]$} such that for some \bl{y}  there is a pair \bl{$[$x, y$]$} in \bl{r}  and a pair \bl{$[$y, z$]$} in \bl{s}\footnote{A commonly used operator for composition is ``\bl{$\circ$}'', which lists the  operands in reverse order. ``\bl{$;$}'' is used among others by the VDM specification language and goes back to  Ernst Schröder in the 19th century$)$ }. $($If \bl{r} $\in$  \bl{A}  \bl{$\leftrightarrow$}   \bl{B} and \bl{s} $\in$  \bl{B}  \bl{$\leftrightarrow$}   \bl{C}, then \bl{r $;$ s} $\in$  \bl{A}  \bl{$\leftrightarrow$}   \bl{C}.$)$

	A relation \bl{f} is a function if it includes, for any given \bl{x}, at most one pair \bl{$[$x, y$]$}. The set of functions from \bl{A} to \bl{B}, a subset of  \bl{A } \bl{\bl{$\leftrightarrow$}} \bl{B}, is written \bl{A $\arrowbar$ B}. If \bl{f} is total $($as a relation, according to the definition  above$)$ then it does have one such pair for every \bl{x} in \bl{A}; we then say it is a total function. The set of total functions  from \bl{A} to \bl{B}, a subset of  \bl{A  $ \arrowbar$ B}, is written \bl{A $\rightarrow$ B}. A function is said to be partial if it is not total\footnote{It is common to use ``partial function'' for what is called just ``function'' here:  a function that is \textit{possibly} partial. The terminology used in this article is more consistent, with ``partial'' meaning the opposite of ``total''.}. For a function \bl{f} and an element \bl{x} of its domain, we may use the function application notation \bl{f $($x$)$} to denote the single element \bl{y} such that \bl{$[$x, y$]$ $\in$ f}\footnote{The image notation from relations remains available, but since the notation is ``typed'' $($it always specifies, for any object in the text, a set of which it is a member$)$ there is no ambiguity: with \bl{f} in \bl{A} $\arrowbar$ \bl{B} and an element \bl{x} of \bl{A}, the function application \bl{f $($x$)$}, if defined, denotes an element of \bl{B}; for a subset \bl{X} of \bl{A}, its image \bl{f $($X$)$}, always defined, denotes a subset of \bl{B}.}. If \bl{f} is total, this notation is always meaningful; otherwise, using it is subject to a \bf{proof obligation} that \bl{x $\in$ {\underline{r}}}.


	\subsection{Lambda notation} \label{lambda}

	Similar to the definition of sets by comprehension as seen above,  ``lambda notation'' is available to define specific relations and functions by their properties. For a boolean-valued property \bl{p $($x, y$)$} applicable to \bl{x} and \bl{y} respectively in \bl{A} and \bl{B}, and an expression \bl{e $($x$)$} denoting an element of \bl{B} for any \bl{x} in \bl{A}, we may use:
	\begin{itemize}
		\item \bl{$\lambda$ x$:$ A $|$ e $($x$)$}$:$ a (total) function in \bl{A} \bl{$\rightarrow$} \bl{B} which for every \bl{x} in \bl{A} yields \bl{e $($x$)$}.

		\item \bl{$\Lambda$ x$:$ A$;$ y$:$ B $|$ p $($x, y$)$}$:$ a relation in \bl{A} \bl{$\leftrightarrow$} \bl{B} which contains the pairs \bl{$[$x, y$]$} such that \bl{p $($x, y$)$} holds. $($It can also be written \bl{\{$[$x, y$]$: A} $\times$ \bl{B} \bl{$|$ p $($x, y$)$\}} but the $\Lambda$ notation emphasizes that we get a relation.$)$

	\end{itemize}

	\noindent Functions as defined so far take a single argument. It can be in a cartesian product: if \bl{f} is in  \bl{A $\times$ B $\rightarrow$ C}, we can apply it to a pair, as in \bl{f $($$[$x, y$]$$)$}, but we can also consider it as two-argument function and write the application as just \bl{f $($x, y$)$}; similarly for more arguments. Lambda notation for functions can similarly use more than one variable.
	The notation also makes it possible to define multi-level functions and relations, which use functions or relations as arguments or results. For example the composition operator ``\bl{$;$}'' can be defined as a function taking two relations and yielding a relation\footnote{In reading this example, make sure not to confuse  the capital ``lambda'' \bl{$\Lambda$} and the boolean ``and'' \bl{$\land$}.}:

	\begin{itemize}
		\item [] \bl{$\lambda$ r: A $\leftrightarrow$}   \bl{B}$;$ \bl{s}$:$ \bl{B}  \bl{$\leftrightarrow$}   \bl{C $|$ $\Lambda$ x: A$;$
			      z: C $|$} \bl{$\exists$} \bl{y$:$ B $|$} \bl{$[$x, y$]$ $\in$ r $\land$ $[$y, z$]$ $\in$ s}

	\end{itemize}

	\noindent Sometimes it is convenient to restrict functions to one argument. One can resort to this convention without any loss of generality by applying a scheme known as \textit{currying}, which from a two-argument function \bl{f} in \bl{A $\times$ B} \bl{$\rightarrow$}   \bl{C} yields a function \bl{ f'}  in \bl{A} \bl{$\rightarrow$}   \bl{$($B} \bl{$\rightarrow$}   \bl{C$)$} such that \bl{$($f'$($x$)$$)$ $($y$)$ = f $($x, y$)$}.


	\subsection{Closures} \label{closures}

	The next notions apply to a relation \bl{r} that is in
	\bl{A } \bl{\bl{$\leftrightarrow$}} \bl{A} for some \bl{A} $($same source and target set$)$.  We may define its powers: \bl{r$^1$} is \bl{r}, \bl{r$^2$} is \bl{r $;$ r} $($\bl{r} composed with itself$)$ and so on\footnote{Formally, ``and so on'' means that \bl{r$^{i+1}$} is defined as \bl{r$^i$ $;$ r}, or equivalently
		as  \bl{r $;$ r$^i$}.}; \bl{r$^0$} is the identity relation \bl{Id $[$A$]$} on \bl{A} $($defined as \bl{$\Lambda$ x$:$ A $|$ $[$x, x$]$} $)$. Another special relation is \bl{Univ $[$A$]$}, defined as \bl{A $\times$ A}, the universal relation, which includes every pair of elements of \bl{A}.  The reflexive transitive closure \bl{r$^*$} of \bl{r} is \bl{ $\cup$} \bl{r$^i$} for all natural integers \bl{i}; excluding from this union of the relations the case \bl{i} = \bl{$0$} yields the $($non-reflexive$)$ transitive closure \bl{r$^+$}.
	A relation \bl{r} is transitive if \bl{r $;$ r $\subseteq$ r}\footnote{In other words, whenever \bl{r} contains two pairs \bl{$[$x, y$]$} and \bl{$[$y, z$]$} it also contains \bl{$[$x, z$]$.}}. $($Equivalently, if \bl{r} = \bl{r$^+$}.$)$ It is symmetric if \bl{r} = \bl{r$^{-1}$} and reflexive if \bl{Id $[$A$]$ $\subseteq$ r}. By construction, the transitive closure $($reflexive or not$)$ of any relation is associative, and the reflexive closure is reflexive. Related concepts for functions of two arguments $($often expressed as operators, such as ``\bl{$;$}'' on relations$)$ are ``associative'' $($\bl{f $($f $($x, y$)$, z$)$ = f $($x, f $($y, z$)$$)$$)$} and ``commutative'' $($\bl{f $($x, y$)$ = f $($y, x$)$$)$}.



	\section{Basic programming concepts} \label{basic}

	\begin{note}{Notational convention}%
		All formal elements $($definitions, theorems...$)$ have a unique identifier such as \referencename{basicdef} below, with a fixed structure $($``slash'' characters bracketing a two-part name with intervening underscore$)$. The same names are used in the publicly available Isabelle/HOL repository \cite{weber2025} associated with this article, which includes the mechanically-verified proofs of all the theorems, enabling any reader to peruse and reproduce them. A reference such as  \referencename{basicdef} is $($in the electronic version$)$ a hyperlink to the defining occurrence.

	\end{note}


	\subsection{Definitions: program, specification, precondition, postcondition} \label{defprogram}

	\begin{colfr}
		\begin{sdef2}{0.80}{0.15}{basicdef}{Basic\_def}
			A \bf{program} $($or \bf{specification}$)$, relative to a set \bl{S}, the \bf{state space}, consists of:
		\end{sdef2}
		\begin{itemize}

			\item \begin{sdef2}{0.80}{0.15}{postcond}{Post\_cond}
				      A relation \bl{post} in \bl{S $\leftrightarrow$ S}, the \bf{postcondition}.
			      \end{sdef2}
			\item \begin{sdef2}{0.75}{0.2}{precond}{Pre\_cond}
				      A subset \bl{Pre} of \bl{S}, the \bf{precondition}.
			      \end{sdef2}
		\end{itemize}
		It may be written \bl{$\langle$post, Pre$\rangle$}.
	\end{colfr}

	\begin{note}{Explanation}%
		A program starts from a certain state and produces one of a set of possible states satisfying properties represented by \bl{post}. \bl{Pre} tells us which states are acceptable as initial. In the general case, more than one resulting state can meet the expectation expressed by \bl{post}. Correspondingly, \bl{post} is a relation rather than just a function. $($See \ref{determinism} below about non-determinism.$)$

	\end{note}
	\begin{note}{Justification}%
		The usual view in software engineering treats “program” and “specification” as distinct concepts, but all definitions of the purported difference are vague: for example, that a specification describes the “what” and a program the “how”, not an meaningful distinction since these notions are relative. An assignment is implementation to the application programmer and specification to the compiler writer. \bf{\textcolor{blue}{out$^2$}} \bl{$\cong$ in} may look like a specification; but some “programming” languages --- and AI tools --- accept it, letting the compiler derive a square-root algorithm. Any useful distinction must be relative: a program/specification “\textit{specifies}” another. The introduction of ``contracted program'' in section \ref{contracts}$)$ will formalize this idea.
	\end{note}

	\begin{note}{Limitation}%
		The definition covers programs/specifications of the input-to-output kind. It can be extended to continuously running programs such as operating systems, and to reactive systems --- extensions that are beyond the scope of the present article.
	\end{note}


	\begin{note}{Notation}%
		\bl{post$_p$}, and \bl{Pre$_p$} $($the latter also written \pdomain{p} per \ref{domainrangeprogram} below$)$ are the postcondition and precondition of a program \bl{p}. If  \bl{p$_i$} is an indexed list of programs, we may use \bl{post$_i$} and \bl{Pre$_i$} as abbreviations for  \bl{post$_{p_i}$} and  \bl{Pre$_{p_i}$}.
	\end{note}

	\begin{note}{Convention}%
		For simplicity, the notations for programs used in this article take the state \bl{S} for granted, assuming that all programs act on the same global \bl{S}. A finer-grain analysis would make the state explicit in each case, adding the notation \bl{S$_p$}, and specify a program with three components as \bl{$\langle$post, Pre, S$\rangle$} rather than just the first two. The definitions and proofs in the Isabelle/HOL complement to this article \cite{weber2025} do take this approach of making the state explicit. A few more observations on the nature of states appear in section \ref{statenature}.

	\end{note}

	\subsection{Definition: deterministic} \label{determinism}

	\begin{colfr}
		A program \bl{p} is:

		\begin{itemize}
			\item \begin{sdef2}{0.6}{0.3}{deterministicdef}{Deterministic\_def}
				      \bf{Deterministic} if \bl{post$_p$} is a function.
			      \end{sdef2}
			\item \begin{sdef2}{0.6}{0.3}{nondeterministicdef}{NonDeterministic\_def}
				      \bf{Non-deterministic} otherwise.
			      \end{sdef2}
		\end{itemize}

	\end{colfr}

	\begin{note}{Explanation}
		For a deterministic program, the postcondition is a function, so that the program always delivers at most one result. $($For an input in \bl{Pre} it delivers \textit{exactly} one result.$)$ In simple sequential programming, programs are usually deterministic; in concurrent and distributed programming, many programs are non-deterministic.

	\end{note}

	\subsection{Definition: domain and range of a program} \label{domainrangeprogram}

	The following notations generalize to a program \bl{p} the domain \bl{\underline{r}} and range \bl{\textoverline{r}} notations for a relation. The first is just an abbreviation for the precondition of a program. The second one denotes the set of states that the program can reach $($if started in its precondition$)$.

	\begin{table}[H]
		\rowcolors{1}{bgcolor}{bgcolor}
		\arrayrulecolor{bordercolor}
		\setlength{\arrayrulewidth}{\borderwidth}
		\setlength{\tabcolsep}{12pt}
		\renewcommand{\arraystretch}{1.8}
		\begin{tabular}{|l| l | l r|}
			\hline
			\bf{Notation}    & \bf{Definition}                            & \multicolumn{2}{l|}{\bf{Name}}                                                    \\
			\hline
			\bl{\pdomain{p}} & \bl{Pre$_p$}                               & Precondition $($``domain''$)$ of a program & \defnraw{prenotation}{Pre\_notation} \\
			\hline
			\bl{\prange{p}}  & \bl{\textoverline{post$_p$ / \pdomain{p}}} & Usable range of a program                  & \defnraw{progrange}{Prog\_range}     \\
			\hline
		\end{tabular}
		\label{tab:domain_range_definition}
	\end{table}

	\noindent From properties of relations $($\referencename{domainrestrict}$)$ it follows, for any set \bl{C} of states, that:
	\begin{colfr}
		\begin{sdef}{0.8}{totalprog2}{Domain\_range}
			\bl{post$_p$ $($\pdomain{p}$)$ =  \prange{p}}
			\\
			\hspace*{2em}       -{}- Notation reminder: left side is image of \bl{\pdomain{p}} by relation \bl{post$_p$.
			}
		\end{sdef}
	\end{colfr}


	\subsection{Definitions: feasible, rounded, exact} \label{feasibility}

	\begin{colfr}
		A program \bl{p} is:
		\begin{itemize}
			\item \begin{sdef2}{0.6}{0.3}{feasibleprog}{Feasible\_prog}\bf{Feasible} if \bl{\pdomain{p} $\subseteq$ \underline{post$_p$}}.\end{sdef2}
			\item \begin{sdef2}{0.6}{0.3}{roundedprog}{Rounded\_prog}\bf{Rounded} if \bl{\underline{post$_p$} $\subseteq$ \pdomain{p}}.\end{sdef2}
			\item \begin{sdef2}{0.6}{0.3}{exactprog}{Exact\_prog}\bf{Exact} if both feasible and rounded.\end{sdef2}
		\end{itemize}
	\end{colfr}

\begin{figure}[hbt!]
		\centering
		\includegraphics[width=0.4 \linewidth]{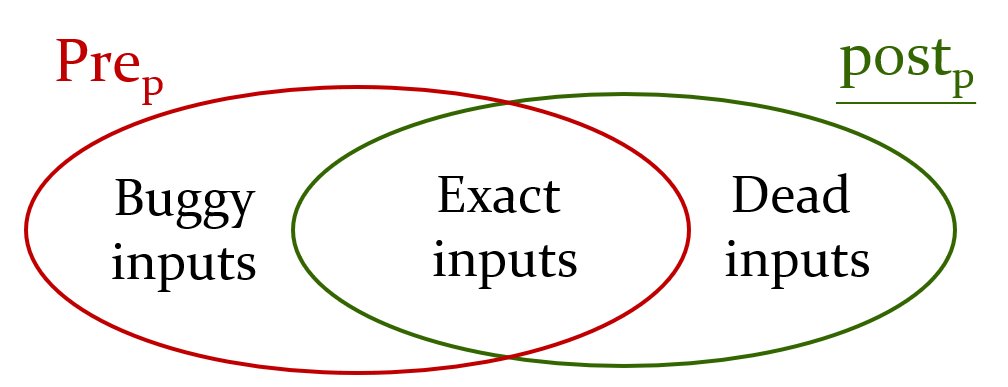}
		\caption{\bf{Categories of input states}}\label{categories}
	\end{figure}

	\begin{note}{Terminology}
		The following terms will denote departures from feasibility and roundedness  $($see Fig. \ref{categories}$)$:
		\begin{itemize}
			\item
			      If a program is \textit{infeasible}, it cannot handle some legal input states $($those in \bl{\pdomain{p} $-$} \bl{\underline{post$_p$}}$)$. We say that the program is \bf{buggy}; such  unhandled cases signal \bf{bugs}.

			\item
			      If the program is \textit{unrounded}, the relation could handle some input states $($those in \bl{\underline{post$_p$} $-$} \bl{\underline{p}}$)$ that are, however, not allowed. We say that the program has \bf{dead code}.

		\end{itemize}

	\end{note}

	\begin{note}{Explanation}
		\bl{\pdomain{p}} tells us when we \textit{may} apply the program, and \bl{post$_p$} what kind of result it \emph{must} then give us. A program/specification is safe for us to use if it meets its obligation whenever we meet ours. Feasibility expresses this property: for any input state satisfying \bl{\pdomain{p}}, at least one output state satisfies \bl{post$_p$}. Roundness, on the other hand, expresses that the relation is used to its full extent: the program will accept all meaningful input states.
	\end{note}

	\begin{note}{Justification}
		\begin{itemize}
			\item  []
			      A program/specification describes a transformation between states, through the relation \bl{post}. The definition, however, also includes a precondition \bl{Pre}. It is legitimate to ask why \bl{Pre} is necessary: would it not be enough to specify the postcondition? Then the precondition would just be \bl{\underline{post}}, the domain of the postcondition.

			\item  []

			      This approach would simplify the theory but is unfortunately not adequate in practice since we are often given a previously-known general relation which may not be applicable to all relevant cases; or, conversely, its application is only relevant for some of the possible cases in its domain. As a simple example, a real-square-root program is defined by the relation  \bl{$\Lambda$ in, out: $\mathbb{R}$ $|$ {out$^2$} $\cong$ in}, a member of \bl{$\mathbb{R}$ $\leftrightarrow$ $\mathbb{R}$}, which is not applicable to every possible input state, but only for a subset of the source set, the set \bl{$\mathbb{R}_{\geq 0}$} of non-negative numbers, which may serve as the precondition.

			\item  []

			      We could in this example consider that the relation is in fact a member of \bl{$\mathbb{R}_{\geq 0}$ $\leftrightarrow$ $\mathbb{R}$}, on which it is total $($and hence does not require a precondition$)$, but even in such a simple case it is inconvenient to have to define endless special input-restricted variants of general relations. In more advanced examples that approach would be even more impractical.

			\item  []

			      The square root example illustrates the need for the precondition to guarantee applicability of the relation; formally, the condition is \bl{\pdomain{p} $\subseteq$ \underline{post$_p$}},  ``feasibility'',  violated  for example if we take \bl{Pre} to be the whole of \bl{$\mathbb{R}$}. We may also encounter the converse problem --- a precondition that is too small $($instead of too large$)$. To avoid this case the program must satisfy the dual condition, roundedness: \bl{\underline{post$_p$} $\subseteq$ \underline{p}}.

		\end{itemize}

	\end{note}

	\begin{note}{Example}
		With \bl{$\mathbb{R}$} as the set of states \bl{S}, consider the square root program where \bl{post} is again \bl{$\Lambda$ in, out: $\mathbb{R}$ $|$ {out$^2$} $\cong$ in}, so that \bl{\underline{post}} is \bl{$\mathbb{R}_{\geq 0}$} since all pairs in the relation have non-negative \bl{in}. Then $($using \bl{$\mathbb{R}^\bullet$} for the set of non-zero numbers and \bl{$\mathbb{R}^{+}$} for the set of positive numbers$)$:
		\begin{itemize}
			\item
			      \bl{$\langle$post, $\mathbb{R}^\bullet$$\rangle$} is not feasible $($it accepts negative numbers but cannot process them$)$ and not rounded $($the relation includes the pair \bl{$[$0, 0$]$} which is not accepted$)$.

			\item \bl{$\langle$post, $\mathbb{R}^+$$\rangle$} is feasible but not rounded $($because of that same pair$)$ .

			\item \bl{$\langle$post $\slash$ $\mathbb{R}_{\geq 0}$, $\mathbb{R}$$\rangle$} is not feasible but is rounded.

			\item \bl{$\langle$post $\slash$ $\mathbb{R}_{\geq 0}$, $\mathbb{R}_{\geq 0}$$\rangle$} is feasible and rounded.

		\end{itemize}
	\end{note}

	\begin{note}{Comment}
		Having to prove feasibility or roundness may seem tedious, but the real culprit is the notion of precondition. In light of the preceding definition, we may surmise that the need to include this notion is in line with general results of incompleteness and undecidability in computer science: we cannot limit ourselves to a comfortable world where all postconditions would be total $($and all programs would terminate$)$. We must accept that many realistic programs will have a non-trivial precondition and that we cannot escape our obligation to satisfy the precondition when applying the program.

	\end{note}











	\subsection{Definitions: trimming and rounding a program} \label{trim}

	\begin{colfr}

		For a given program \bl{p}:
		\begin{itemize}

			\item \begin{sdef2}{0.6}{0.3}{feasibleversion}{Feasible\_version}\bl{$\lceil{p}\rceil$}, the \bf{feasible version} of \bl{p}, is \bl{$\langle$post$_p$, \pdomain{p} $\cap$ \underline{post$_p$}$\rangle$}.\end{sdef2}

			\item \begin{sdef2}{0.6}{0.3}{roundedversion}{Rounded\_version}\bl{$\lfloor{p}\rfloor$}, the \bf{rounded} \bf{version} of \bl{p}, is \bl{$\langle$post$_p$ / \pdomain{p}, \pdomain{p}$\rangle$}.\end{sdef2}

			\item \begin{sdef2}{0.7}{0.2}{exactversion}{Exact\_version}\bl{$[{p}]$}, the \bf{exact version} of \bl{p}, is \bl{$\langle$post$_p$  / \pdomain{p}, \pdomain{p} $\cap$ \underline{post$_p$}$\rangle$}.\end{sdef2}
		\end{itemize}
	\end{colfr}

	\begin{note}{Terminology}
		The feasible version is also called the ``trimmed'' version. ``Making a program feasible'', or synonymously ``trimming'' it, means replacing it by its feasible version. ``Rounding'' it means replacing it by its rounded version.
	\end{note}

	\begin{note}{Explanation}
		In line with the earlier explanations $($section \ref{feasibility}$)$, we may view \bl{$\lceil{p}\rceil$} as \bl{p} cleaned of dead code, and \bl{$\lfloor{p}\rfloor$} as \bl{p} freed from bugs, in other words restricted to input states that it can actually process.
	\end{note}

	\begin{note}{Example}
		For  square root, with \bl{$\mathbb{R}$} as the set of states \bl{S} and \bl{p $=$ $\langle$post, $\mathbb{R}^\bullet\rangle$} where \bl{post} is \bl{$\Lambda$ in, out: $\mathbb{R}$ $|$ {out$^2$} $\cong$ in}, so that \bl{p}, as seen above, is neither feasible nor rounded:
		\begin{itemize}
			\item
			      \bl{$\lceil{p}\rceil$} is \bl{$\langle$post, $\mathbb{R}^{+}\rangle$} $($rejecting ``buggy'' negative inputs, but retaining ``dead code'' in the postcondition$)$, feasible and not rounded.

			\item \bl{$\lfloor{p}\rfloor$} is \bl{$\langle$post $\slash$ $\mathbb{R}^{\bullet}$, $\mathbb{R}\rangle$} $($eliminating dead code but retaining buggy input$)$, rounded and not feasible.

			\item \bl{$[{p}]$} is \bl{$\langle$post $\slash$ $\mathbb{R}^{\bullet}$, $\mathbb{R}^{+}\rangle$} $($without dead code or bugs$)$, exact. The postcondition could also be \bl{post $\slash$ $\mathbb{R}^{+}$}.
		\end{itemize}
	\end{note}

	\begin{colfr}[Feasibility and roundness theorems]
		For any program \bl{p}:\\\\
		\vspace{0.2cm}\begin{sdef}{0.7}{consistentfeasible}{Consistent\_feasible}
			\hspace{2em} \bl{$\lceil$p$\rceil$} is feasible. \\
			\hspace*{5em}   -{}- {The ``feasible version'' deserves its name.}
		\end{sdef}
		\vspace{0.2cm}\begin{sdef}{0.7}{consistentround}{Consistent\_round}
			\hspace{2em} \bl{$\lfloor$p$\rfloor$} is rounded. \\
			\hspace*{5em}   -{}- {The ``rounded version'' deserves its name.}
		\end{sdef}
		\vspace{0.2cm}\begin{sdef}{0.7}{consistentexact}{Consistent\_exact}
			\hspace{2em} \bl{$[p]$} is exact. \\
			\hspace*{5em}   -{}- {The ``exact version'' deserves its name.}
		\end{sdef}
		\vspace{0.2cm}\begin{sdef}{0.7}{feasibleround}{Feasible\_round}
			\hspace{2em} If \bl{p} is feasible, so is \bl{$\lfloor$p$\rfloor$}. \\
			\hspace*{5em}   -{}- {Rounding preserves feasibility.}
		\end{sdef}
		\vspace{0.2cm}\begin{sdef}{0.7}{roundfeasible}{Round\_feasible}
			\hspace{2em} If \bl{p} is rounded, so is \bl{$\lceil$p$\rceil$}. \\
			\hspace*{5em}   -{}- {Trimming $($making feasible$)$ preserves rounding.}
		\end{sdef}

	\end{colfr}
	\begin{note}{Proof idea}%
		\referencename{feasibleround} follows from a property of restriction: if \bl{A $\subseteq$ \underline{r}}, then \bl{A $\subseteq$ \underline{r / A}}, applied to the relation \bl{post$_p$} and the set \bl{\pdomain{p}}. \referencename{roundfeasible} follows from the elementary set-theoretical property that if \bl{A $\subseteq$ B}, then \bl{A $\subseteq$ A $\cap$ B}, applied to the sets \bl{\underline{post$_p$}} and \bl{\pdomain{p}}.
	\end{note}

	\begin{note}{Notations}
		\begin{itemize}
			\item
			      \bl{post$_p$ / \pdomain{p}}, the postcondition of  \bl{$\lfloor p\rfloor$}, may be written \bl{$\lfloor$post$_p\rfloor$}.
			\item
			      \bl{\pdomain{p} $\cap$ \underline{post$_p$}}, the precondition of  \bl{$\lceil p\rceil$}, may be written \bl{$\lceil$\pdomain{p}$\rceil$}.

		\end{itemize}
	\end{note}


	\subsection{Definition: total} \label{total}
	\begin{colfr}
		\begin{sdef}{0.8}{totalprogram}{Total\_program}
			A program \bl{p} is \bf{total} if it is feasible and \bl{\pdomain{p} = S}.

		\end{sdef}
	\end{colfr}

	\begin{note}{Explanation}
		A total program can process all states. In informal parlance, it ``has no precondition''. $($Formally, it does have a precondition: the entire state set \bl{S}, also called \bl{True} per \ref{conditions} below.$)$
	\end{note}

	\begin{note}{Comment}
		This definition of ``total'' for a program is consistent with the definition of ``total'' for a relation  in section \ref{notation} in the following sense: \bl{p} is a total program if and only if it is feasible and \bl{post$_p$} is a total relation.
	\end{note}


	\subsection{Definitions: program equality and equivalence} \label{equality}

	\begin{colfr}
		Two programs \bl{p} and \bl{q} are:
		\begin{itemize}
			\item \begin{sdef2}{0.75}{0.2}{equalprogram}{Equal\_program}\bf{Equal}, written \bl{p = q}, if  they have the same postconditions and preconditions $($that is, \bl{$post_p$ = $post_q$} and   \bl{\pdomain{p} = \pdomain{q}}$)$.\end{sdef2}
			\item \begin{sdef2}{0.75}{0.2}{equivprogram}{Equiv\_program}\bf{Equivalent}, written \bl{p $\equiv$ q}, if their rounded versions are equal $($that is,   \bl{$\lfloor p \rfloor$ = $\lfloor q \rfloor$}$)$.\end{sdef2}
		\end{itemize}
	\end{colfr}

	\begin{note}{Comments}
		The first definition is the usual notion of equality $($since a program is defined by its postcondition and precondition$)$. As elsewhere in this article, we assume that the state set \bl{S} is the same for all programs.
	\end{note}

	\begin{note}{Justification}
		The theorems introduced in the following sections sometimes state that two expressions yield equal programs, but sometimes one or both of the programs have some extra pairs $($``dead code''$)$ in their \bl{post}. In such cases only the rounded versions are equal and the theorems assert equivalence rather than full equality.
	\end{note}

	\begin{note}{Property}
		Obviously, equivalence $($like any relation of the form ``\bl{p} and \bl{q} have the same <something>''$)$ is an equivalence relation in the usual sense $($reflexive, symmetric, transitive$)$.
	\end{note}


	\subsection{Conditions} \label{conditions}

	\begin{note}{Terminology}
		In the rest of this article, the term ``condition'' will denote a subset $($usually called \bl{C}, \bl{D} or \bl{Pre}$)$ of the set of states.  \bl{S}.  A precondition is a condition, but a postcondition is not $($it is a relation on states$)$. The following notations from logic can be used in lieu of their set-theoretical counterparts.
	\end{note}

	\begin{table}[H]
		\rowcolors{1}{bgcolor}{bgcolor}
		\arrayrulecolor{bordercolor}
		\setlength{\arrayrulewidth}{\borderwidth}
		\setlength{\tabcolsep}{12pt}
		\renewcommand{\arraystretch}{1.8}
		\begin{tabular}{|l| l r|}
			\hline
			\bf{Notation}          & \multicolumn{2}{l|}{\bf{Definition}}                                              \\
			\hline
			\bl{True}              & \bl{S}                               & \defnraw{truedef}{True\_def}               \\
			\hline
			\bl{False}             & \bl{$\varnothing$}                   & \defnraw{falsedef}{False\_def}             \\
			\hline
			\bl{C \bf{$\lor$} D}   & \bl{C $\cup$ D}                      & \defnraw{ordef}{Or\_def}                   \\
			\hline
			\bl{C \bf{$\land$} D}  & \bl{C $\cap$ D}                      & \defnraw{anddef}{And\_def}                 \\
			\hline
			\bl{\bf{$\lnot$} C}    & \bl{C = False}                       & \defnraw{notdef}{Not\_def}                 \\
			\hline
			\bl{C $\Rightarrow$ D} & \bl{C $\subseteq$ D}                 & \defnraw{impliesarrow}{Implies\_arrow}     \\
			\hline
			\bl{C \bf{implies} D}  & \bl{C $\subseteq$ D}                 & \defnraw{implieskeyword}{Implies\_keyword} \\

			\hline
		\end{tabular}
		\label{tab:boolean_definition}
	\end{table}

	\begin{note}{Comment}
		\bl{C = True} is equal to \bl{C}. The rest of this article uses the notations defined here except for ``\bl{$\lor$}'' and ``\bl{$\land$}'' for which the set operators are just as convenient.
	\end{note}

	\begin{bcolfr} [Theorems: properties of special conditions]
		For any program \bl{p}:\\
		\begin{sdef}{0.8}{restricttrue}{Restrict\_true}
			\hspace*{3em}\bl{True: p = p}
		\end{sdef}
	\end{bcolfr}
	\begin{bcolfr}
		\begin{sdef}{0.8}{restrictfalse}{Restrict\_false}
			\hspace*{3em}\bl{False: p $\equiv$ Fail}
		\end{sdef}
	\end{bcolfr}
	\begin{bcolfr}
		\begin{sdef}{0.8}{corestricttrue}{Corestrict\_true}
			\hspace*{3em}\bl{p $\backslash$ True $\equiv$ p} for feasible  \pr.
		\end{sdef}
	\end{bcolfr}
	\begin{bcolfr} [bot]
		\begin{sdef}{0.8}{corestrictfalse}{Corestrict\_false}
			\hspace*{3em}\bl{p $\backslash$ False = Fail}
		\end{sdef}
	\end{bcolfr}


	\subsection{The nature of states} \label{statenature}

	\noindent Other than in specific illustrative examples, this article ignores what the state set \bl{S} actually is; it is indeed possible to obtain many properties of programs and programming without considering the structure of \bl{S}. That is what the following sections do.

	In other contexts $($and beyond the scope of the present article$)$ it becomes necessary to look into specific kinds of state, which yield specific forms of programming. Here we may simply note that useful forms of states generally include a \textit{binding}, defining the relationship between elements of programs and elements of their executions. A binding \bl{b} is a function in \bl{Name $\arrowbar$ Value} for some sets \bl{Name} $($related to a program's text$)$ and \bl{Value} $($related to the program's executions$)$.Two special kinds of program are particularly interesting:

	\begin{colfr}
		A program \bl{p} is:
		\begin{itemize}
			\item \begin{sdef2}{0.7}{0.2}{functionaldef}{Functional\_def}
				      \bf{Functional} if \bl{b $\subseteq$ $($post $;$ b$)$}.
			      \end{sdef2}
			\item \begin{sdef2}{0.8}{0.15}{oodef}{OO\_def}
				      \bf{Object-oriented} if \bl{Name} is \bl{1}..\bl{n} for some integer \bl{n}. The elements of \bl{Value} in this case are called ``objects'' and the binding is also called a ``heap''.
			      \end{sdef2}
		\end{itemize}
	\end{colfr}

	\begin{note}{Explanation}

		\begin{itemize}
			\item A functional program does not modify any existing name-value association. In other words, the binding that holds in a state \bl{s} remains in any of the states in \bl{post $($s$)$}: for all \bl{s}, \bl{b $($s$)$} $\subseteq$ \bl{b $($post $($s$)$$)$} as expressed by the definition. For the binding and the order relation defined by \bl{$\subseteq$}, the postcondition is non-decreasing. Execution  can enrich the binding with new \bl{$[$name, value$]$} pairs but never remove any pair.
			\item An object-oriented program manipulates a set of values each with a unique ``object id''.
		\end{itemize}
		The two definitions are unrelated, with the consequence that a program can be both functional and object-oriented.
		The fundamental program constructs and properties reviewed in the rest of this article are entirely independent of the nature of state and the resulting programming style $($such as functional or object-oriented$)$.

	\end{note}

	\vspace{-0.5cm}
	\section{Fundamental program constructs} \label{constructs}

	The following operations on programs are the basic building blocks of programming. They are not typically used directly as programming language constructs $($although they could be$)$, but serve as a basis for the actual language constructs introduced from sections \ref{conditionals} on.

	The present section also introduces a set of extreme-case programs with simple definitions and distinctive properties, similar to special values $($zero, unity$)$ in other mathematical theories.


	\subsection{Overview of the basic operators} \label{operatoroverview}

	There are three basic constructs. Informally $($for programs \bl{p, q} and a set of states \bl{C}$)$:
	\begin{itemize}
		\item \bl{C$:$ p}, the \bf{restriction} of \bl{p} to \bl{C} executes like \bl{p}, but is applicable only in \bl{C}.

		\item \bl{p $\cup$ q}, the \bf{choice} $($or union$)$  between \bl{p} and \bl{q}, executes like either \bl{p} or \bl{q} whenever one of them is applicable. $($If both are, the execution could use either one.$)$

		\item \bl{p $;$ q}, the \bf{composition} of \bl{p}  and  \bl{q}, executes first like \bl{p} then like \bl{q}.
	\end{itemize}

	\begin{note}{Notation reminder}
		\bl{r $\slash$ C} and \bl{r $\backslash$ D} are the restriction and corestriction of a relation \bl{r} to subsets \bl{C} of its source set and \bl{D} of its target set.
	\end{note}


	\subsection{Definitions: restriction, choice and composition} \label{basicops}

	\begin{table}[H]
		\rowcolors{1}{bgcolor}{bgcolor}
		\arrayrulecolor{bordercolor}
		\setlength{\arrayrulewidth}{\borderwidth}
		\setlength{\tabcolsep}{12pt}
		\renewcommand{\arraystretch}{1.8}
		\begin{tabular}{|c|c|c|c r|}
			\hline
			\bf{Notation}   & \bf{Name}    & \bf{Postcondition}                           & \multicolumn{2}{l|}{\bf{Precondition}}                                                               \\

			\hline
			\bl{C$:$ p}     & Restriction  & \bl{post $\slash$ C}                         & \bl{Pre $\cap$ C}                                             & \defnraw{restrictdef}{Restrict\_def} \\

			\hline
			\bl{p $\cup$ q} & \shortstack{                                                                                                                                                       \\ Choice \\ $($or: union$)$} & \bl{$\lfloor$post$_p$$\rfloor$ $\cup$ $\lfloor$post$_q$$\rfloor$ } & \bl{\pdomain{p} $\cup$ \pdomain{q} }& \defnraw{choicedef}{Choice\_def} \\

			\hline
			\bl{p $;$ q}    & Composition  & \bl{\post[p] $;$ $\lfloor$\post[q]$\rfloor$} & \bl{\Pre[p] $\cap$ \underline{\post[p] $\backslash$ \Pre[q]}} & \defnraw{composedef}{Compose\_def}   \\

			\hline
		\end{tabular}
		\label{tab:program_operation}
	\end{table}
	\begin{note}{Explanation}%
		See the informal introduction above.
	\end{note}

	\begin{note}{Justification}
		These definitions embody the core thesis of this work, considering that program operations are essentially applications of classical operations from elementary set theory:
		\begin{itemize}
			\item Restriction on programs is restriction of the postcondition and precondition.
			\item Union of programs is union of the $($rounded$)$ postconditions and precondition.
			\item Composition on programs is composition of the postconditions, with the necessary narrowing of the precondition and again rounding.
		\end{itemize}

	\end{note}

	\begin{note}{Notation}
		The operators for programs reuse the symbols of the corresponding set operators. ``\bl{ $\cup$ }'' and ``\bl{;}'' in the first column above are union and composition of programs; in the other columns, they are union of sets and composition of relations. Restriction for programs could be written \bl{p $\backslash$ C} like its relation counterpart; the semicolon notation reflects programmers' practice of writing the condition first, as in ``if \bl{C} holds, execute \bl{p}''. $($The actual ``conditional instruction'', based on restriction and choice, appears below in section \ref{conditionals}.$)$

	\end{note}
	\begin{note}{Comment}%
		The rounding of postconditions is necessary for consistency:
	\end{note}
	\begin{itemize}
		\item
		      For choice, using only \bl{post$_p$ $\cup$ post$_q$} without rounding may lead $($if the original programs are not rounded$)$ to wrongly resuscitating dead code; specifically, applying \bl{post$_q$} in a state in \bl{\pdomain{q} -- \pdomain{p}} $($or conversely$)$. For example, if \bl{p} is \bl{$\langle\{[1, 1], [2, 1]\}, \{1\}\rangle$} and \bl{q} is \bl{$\langle\{[2, 2]\}, \{2\}\rangle$} then the union of the postconditions --- without rounding --- includes \bl{$[2, 1]$}, which is applicable under the union of the preconditions \bl{$\{1, 2\}$}, even though it is applicable under neither of the original programs. The definition of choice removes such anomalies by rounding the two operands.

		\item Similarly for composition, a dead pair \bl{$[$b, c$]$} of \bl{post$_q$} $($dead because \bl{b} is not in \bl{\pdomain{q}}$)$ might find its match in a pair \bl{$[a, b]$} of \bl{post$_p$}, bringing it back to life. For example, if \bl{p} is \bl{$\langle\{[1, 10], [1, 20]\}, \{1\}\rangle$} and \bl{q} is \bl{$\langle\{[10, 10], [20, 1000]\}, \{10\}\rangle$}, with the second pair of its postcondition dead, then the composition of the postconditions --- without rounding of the second one --- will take advantage of that pair to include, along with the expected pair \bl{$[1, 10]$}, the spurious pair \bl{$[1, 1000]$} in the result. The definition of composition removes such anomalies by rounding the second operand.
	\end{itemize}


	\begin{note}{Notational variants}%
		\begin{itemize}
			\item

			      Choice might also be called union and composition sequence, compound or block.

			\item
			      It would also be possible to use Dijkstra's symbols for ``guarded commands'' \cite{dijkstra1975}: \bl{p $[$\hspace{-0.5pt}$]$ q} for \bl{p $\cup$ q} and \bl{C $\rightarrow$ p} for \bl{C$:$ p}.

		\end{itemize}
	\end{note}

	\begin{note}{Comment}%
		In a version of the theory that makes the state explicit, the respective state sets resulting from the three basic operations are \bl{S$_p$}, \bl{S$_1$ $\cup$ S$_2$} and again  \bl{S$_1$ $\cup$ S$_2$}.
	\end{note}


	\subsection{Basic programs} \label{special}

	\begin{note}{Notation reminder}%
		\bl{True} is another notation for \bl{S}, the set of states, and \bl{False} for the empty set of states \bl{$\varnothing$}. \bl{Id [X]} is the identity relation and \bl{Univ [X]}in \bl{X $\leftrightarrow$ X}.  \bl{$\langle post, Pre \rangle$} defines a program by its postcondition and precondition, in that order; \bl{False}  denotes the empty relation in the first position $($where it may also be written \bl{$\varnothing$}$)$, and \bl{$\varnothing$} denotes an empty  relation in the second position.
	\end{note}

	\begin{table}[H]
		\rowcolors{1}{bgcolor}{bgcolor}
		\arrayrulecolor{bordercolor}
		\setlength{\arrayrulewidth}{\borderwidth}
		\setlength{\tabcolsep}{12pt}
		\renewcommand{\arraystretch}{1.8}
		\begin{tabular}{|c| c r|}
			\hline
			\bf{Name}                                    & \multicolumn{2}{l|}{\bf{Definition}}                                              \\

			\hline

			\bl{Fail}                                    & \bl{$\langle \varnothing, ~ \pwrong \rangle$} & \defnraw{faildef}{Fail\_def}      \\
			\hline

			\bl{Infeasible}                              & \bl{$\langle \varnothing, ~ \pright \rangle$} & \defnraw{infeas}{Infeasible\_def} \\
			\hline

			\bl{Havoc}                                   & \bl{$\langle Univ ~ [S], ~ \pright \rangle$}  & \defnraw{havocdef}{Havoc\_def}    \\
			\hline

			\bl{Skip}                                    & \bl{$\langle$Id [S], ~ \pright $\rangle$}     & \defnraw{skipdef}{Skip\_def}      \\
			\hline

			\bl{Skip$_C$} $($for \bl{C $\subseteq$ S}$)$ & \bl{$\langle$Id [C], ~ C $\rangle$}           & \defnraw{skiprestr}{Skip\_restr}  \\
			\hline

			\hline
		\end{tabular}
		\label{tab:basic_programs}
	\end{table}

	\begin{note}{Explanation}%
		\begin{itemize}
			\item \bl{Fail} is never applicable --- and hence never produces any result. This program is sometimes also known as ``abort''.

			\item \bl{Infeasible} is more devious: it accepts any input but  never produces any result. Compare with \bl{Fail} which at least tells you right away that something is wrong: its execution can never proceed. \bl{Infeasible} pretends to proceed but after that nothing more can happen.

			\item \bl{Havoc} accepts any input but does not guarantee anything about the resulting state. Something happens but \textit{anything} can happen.

			\item \bl{Skip} is always applicable but yields a change identical to what it was. Such a program is also called a ``no-op'' $($short for ``no operation''$)$.

			\item \bl{Skip$_C$} is like \bl{Skip} but applicable only on a specific set of states \bl{C}. Clearly, \bl{Skip} is the same as \bl{Skip$_C$}.

		\end{itemize}

		\noindent \bl{Fail}, \bl{Havoc} and \bl{Skip} are all feasible programs. \bl{Infeasible} is not.

	\end{note}

	\begin{note}{Comment}
		These definitions, other than for \bl{Skip$_C$}, assume a single set of states \bl{S}. A more fine-grain definition is also possible, making the state set explicit and indexing all the basic programs with it, as with \bl{Skip$_C$}.
	\end{note}

	\begin{note}{Properties}
		It will follow from the theorems below that the set of programs is a monoid $($associative, identity element$)$ for ``\bl{$;$}''  and \bl{Skip}, and a semilattice $($monoid, commutative, idempotent$)$  for ``\bl{ $\cup$ }'' and \bl{Fail}.
	\end{note}

	\subsection{Corestriction} \label{corestriction}

	Symmetrically with restriction on programs $($\ref{basicops}, derived from restriction on relations \ref{relations}, we may define a corestriction for programs out of corestriction for relations. This operator is less fundamental than the three basic ones studied earlier, but does have its uses.

	\begin{table}[H]
		\rowcolors{1}{bgcolor}{bgcolor}
		\arrayrulecolor{bordercolor}
		\setlength{\arrayrulewidth}{\borderwidth}
		\setlength{\tabcolsep}{4pt}
		\renewcommand{\arraystretch}{1.2}
		\begin{tabular}{|c |c|l r|}
			\hline
			\bf{Notation}         & \bf{Name}     & \multicolumn{2}{l|}{\bf{Definition}}                                                                   \\
			\hline
			\bl{p $\backslash$ D} & Corestriction & \bl{$\langle$\post[p] $\backslash$ D, \pdomain{p}$\rangle$} & \defnraw{corestrictdef}{Corestrict\_def} \\


			\hline
		\end{tabular}
		\label{tab:corestriction}
	\end{table}

	\begin{note}{Explanation}
		\bl{D} is a set of states. \bl{p $\backslash$ D}  the \bf{corestriction} of \bl{p} to \bl{D}. executes like \bl{p}, but only keeps results that belong to \bl{D}.
	\end{note}

	\begin{note}{Caveat}
		A feasible program may become infeasible through corestriction, since some initial states satisfying \bl{\pdomain{p}} may yield final states not satisfying \bl{D}. To guarantee feasibility, use \bl{$\lceil$p $\backslash$ D$\rceil$}. An alternative definition of corestriction, guaranteeing feasibility of the result, would use $($instead of  \pdomain{p}$)$ the precondition \bl{\pdomain{p} $\cap$   \underline{\post[p] $\backslash$ C}}. See \referencename{composeprepost} in \ref{corestrictionproperties} below.
	\end{note}



	\section{Properties of basic programs and operators} \label{basicproperties}

	The basic programs and fundamental operators introduced above satisfy important properties, some of which are listed below. As other theorems in this article, these properties have been proved with Isabelle/HOL \cite{isabelle}. To facilitate connecting the proofs and the present article, the  names appearing here for every property, such as \bl{Skip\_compleft}, also serve as labels of the corresponding proofs in the Isabelle files.


	\subsection{Properties of basic programs} \label{specialprops}

	\begin{colfr}[Theorems: Properties of \bl{Skip}]
		For a program \bl{p} and a subset \bl{C} of \bl{S}:\par

		\vspace{0.2cm}\begin{sdef}{0.8}{skipcomp1}{Skip\_compleft}
			\hspace{2em}\bl{p $;$ Skip = \pr} \\
			\hspace*{5em}   -{}- {\bl{Skip}  is a neutral element for composition, left...}
		\end{sdef}

		\vspace{0.2cm}\begin{sdef}{0.8}{skipcompright}{Skip\_compright}
			\hspace{2em}\bl{Skip $;$ p $\equiv$ \pr} \\
			\hspace*{5em}   -{}- {... and right.}
		\end{sdef}

		\vspace{0.2cm}\begin{sdef}{0.8}{skipempty}{Skip\_empty}
			\hspace{2em}\bl{Skip$_{\pwrong}$ $=$ Fail} \\
			\hspace*{5em}   -{}- {Restriction to nothing is failure.}
		\end{sdef}

		\vspace{0.2cm}\begin{sdef}{0.8}{skipcomprestrict}{Skip\_comprestrict}
			\hspace{2em}\bl{Skip$_C$ $;$ p $\equiv$ C$:$ p} \\
			\hspace*{5em}   -{}- {Left-composing with restricted \bl{Skip} yields restriction...}
		\end{sdef}

		\vspace{0.2cm}\begin{sdef}{0.7}{skipcomposecorestrict}{Skip\_composecorestrict}
			\hspace{2em}\bl{p $;$ Skip$_C$  $\equiv$ p $\backslash$ C} \\
			\hspace*{5em}   -{}- ... and right-composing yields corestriction.
		\end{sdef}

	\end{colfr}

	\begin{note}{Comment}
		In contrast with its properties for composition a, \bl{Skip} obeys no general property vis-à-vis choice. \bl{Skip $\cup$ Skip = Skip} is a consequence of more general properties of choice (see \referencename{choiceidem} in \ref{choiceproperties}).

	\end{note}

	\begin{bcolfr}[Theorems: Properties of \bl{Fail} and   \bl{Havoc}]
		For feasible \bl{p}:\par

	\end{bcolfr}
	\begin{bcolfr}
		\begin{sdef}{0.8}{failchoice}{Fail\_choice}
			\hspace{2em}\bl{$($Fail $\cup$ \pr$)$ $\equiv$ $($\pr $\cup$ Fail$)$ $\equiv$ \pr}
			\par\setlength{\leftskip}{4em}
			\hspace*{5em}-{}- {\bl{Fail} is a neutral element of choice, left and right.}
		\end{sdef}
	\end{bcolfr}
	\begin{bcolfr}
		\begin{sdef}{0.8}{failcomp}{Fail\_comp}
			\hspace{2em}\bl{$($Fail $;$ \pr$)$ = $($\pr $;$ Fail$)$} = \bl{Fail}
			\par\setlength{\leftskip}{4em}
			\hspace*{5em}   -{}- {\bl{Fail} absorbs composition, left and right.}
		\end{sdef}
	\end{bcolfr}
	\begin{bcolfr}
		\begin{sdef}{0.75}{failchoiceonly}{Fail\_choiceonly}
			\hspace{2em}\bl{$($a $\cup$ b$)$ $\equiv$ Fail} if and only if \bl{a $\equiv$ Fail} and \bl{b $\equiv$ Fail}
			\par\setlength{\leftskip}{4em}
			\hspace*{5em}   -{}- {Choice can only  fail if both operands fail.}
		\end{sdef}
	\end{bcolfr}
	\begin{bcolfr}[bot]
		\begin{sdef}{0.8}{havoc_choice}{Havoc\_choice}
			\hspace{2em}$($\pr \bl{ $\cup$ } \bl{Havoc$)$ = \bl{$($Havoc $\cup$ p$)$} = Havoc}
			\par\setlength{\leftskip}{4em}
			\hspace*{5em}   -{}- {\bl{Havoc} absorbs union.}
		\end{sdef}
	\end{bcolfr}



	\subsection{Properties of restriction} \label{restrictionproperties}

	\begin{bcolfr}[Theorems: Properties of restriction]
		For programs \bl{p}, \bl{q} and conditions \bl{C}, \bl{D}:\\\\
		\begin{sdef}{0.8}{restrictown}{Restrict\_own}
			\hspace{2em}\bl{$($\pdomain{p}: p$)$ = p} \\
			\hspace*{5em}   -{}- {Restriction to your own precondition is no restriction.}
		\end{sdef}
	\end{bcolfr}
	\begin{bcolfr}
		\begin{sdef}{0.8}{restrictinter}{Restrict\_inter}
			\hspace{2em}\bl{C$:$ $($D: p$)$ = $($\bl{C} $\cap$ \bl{D}$)$: p} \\
			\hspace*{5em}   -{}- {Repeated restriction is intersection.}
		\end{sdef}
	\end{bcolfr}
	\begin{bcolfr}
		\begin{sdef}{0.8}{restrictcommute}{Restrict\_commute}
			\hspace{2em}\bl{C$:$ $($\bl{D: p$)$ = D: $($C}: p$)$}
			\\ \hspace*{5em}   -{}- {As a consequence, restriction is commutative.}
		\end{sdef}
	\end{bcolfr}
	\begin{bcolfr}
		\begin{sdef}{0.8}{restrictidem}{Restrict\_idem}
			\hspace{2em}\bl{C$:$ $($C$:$ p$)$ = C$:$ p$)$}
			\\ \hspace*{5em}   -{}- {Another consequence is that it is idempotent.}
		\end{sdef}
	\end{bcolfr}
	\begin{bcolfr}
		\begin{sdef}{0.8}{restrictdistrib}{Restrict\_distrib}
			\hspace{2em}\bl{C$:$ $($p $\cup$ q$)$ $\equiv$ C$:$ p $\cup$ C$:$ q}
			\\ \hspace*{5em}   -{}- {Restriction distributes over choice. {Equivalence only: }
			\\ \hspace*{5em}   -{}- restriction may introduce dead code that choice removes.
			\\ \hspace*{5em}   -{}- Example: \bl{p = q = $\langle\{[1,1]\}, \{1\}\rangle$}, \bl{C = $\pwrong$.}}
		\end{sdef}
	\end{bcolfr}
	\begin{bcolfr}[bot]
		\begin{sdef}{0.8}{composeabsorb}{Compose\_absorb}
			\hspace{2em}\bl{C$:$ $($p $;$ q$)$} = \bl{C$:$ p $;$ q}
			\\
			\hspace*{5em}   -{}- {Composition absorbs restriction.}
		\end{sdef}
	\end{bcolfr}

	\subsection{Properties of corestriction} \label{corestrictionproperties}

	\begin{bcolfr}[Theorems: Properties of corestriction]
		For a program \pr and conditions \bl{C} and \bl{D}:\\\\
		\begin{sdef}{0.8}{corestrictinter}{Corestrict\_inter}
			\hspace{2em}\bl{$($\pr $\backslash$C$)$ $\backslash$ D} = \pr $\backslash$ $($\bl{C} $\cap$ \bl{D}$)$
			\\ \hspace*{5em}   -{}- {Repeated corestriction is intersection.}
		\end{sdef}
	\end{bcolfr}
	\begin{bcolfr}
		\begin{sdef}{0.7}{corestrictcommute}{Corestrict\_commute}
			\hspace{2em}\bl{$($\pr $\backslash$ C$)$ $\backslash$ D = \pr $\backslash$\bl {D} $\backslash$ $($C$)$}$)$
			\\ \hspace*{5em}   -{}- {As a consequence, corestriction is commutative.}
		\end{sdef}
	\end{bcolfr}
	\begin{bcolfr}
		\begin{sdef}{0.8}{corestrictidem}{Corestrict\_idem}
			\hspace{2em}\bl{C$:$ $($C$:$ \pr$)$ = C$:$ \pr}
			\\ \hspace*{5em}   -{}- {Another consequence is that it is idempotent.}
		\end{sdef}
	\end{bcolfr}
	\begin{bcolfr}
		\begin{sdef}{0.8}{corestrictchoice}{Corestrict\_choice}
			\hspace{2em}\bl{$($p $\cup$ q$)$ $\backslash$C= \bl{p$\backslash$C $\cup$ q$\backslash$C}}
			\\
			\hspace*{5em}   -{}- {Corestriction distributes over choice.}
		\end{sdef}
	\end{bcolfr}
	\begin{bcolfr}
		\begin{sdef}{0.8}{corestrictcompose}{Corestrict\_compose}
			\hspace{2em}\bl{$($p $;$ q$)$ $\backslash$ C = p $;$ $($q $\backslash$ C$)$}
			\\
			\hspace*{5em}   -{}- {Corestriction is only applied to the last element.}
		\end{sdef}
	\end{bcolfr}
	\begin{bcolfr}[bot]
		\begin{sdef}{0.8}{composeprepost}{Compose\_prepost}
			\hspace{2em}If \pr is feasible then \bl{$($$($\pdomain{p} $\cap$   \underline{\post[p] $\backslash$ C}$)$: p$)$ $\backslash$ C} is feasible.
		\end{sdef}
	\end{bcolfr}


	\subsection{Properties of choice} \label{choiceproperties}

	\begin{bcolfr}[Theorems: Properties of choice]
		For programs \bl{p}, \bl{q}, \bl{v}:\\\\
		\begin{sdef}{0.8}{choicecommute}{Choice\_commute}
			\hspace{2em}\bl{p $\cup$ q = q $\cup$ p} \\
			\hspace*{5em}   -{}- {Choice is commutative.}
		\end{sdef}
	\end{bcolfr}
	\begin{bcolfr}
		\begin{sdef}{0.8}{choiceassoc}{Choice\_assoc}
			\hspace{2em}\bl{p $\cup$ $($q $\cup$ v$)$ = $($p $\cup$ q$)$ $\cup$ v} \\
			\hspace*{5em}   -{}- {Choice is associative.}
		\end{sdef}
	\end{bcolfr}
	\begin{bcolfr}
		\begin{sdef}{0.8}{choiceidem}{Choice\_idem}
			\hspace{2em}\bl{p $\cup$ p $\equiv$ p} \\ \hspace*{5em}   -{}- {Choice is idempotent $($equivalence only, not equality$)$} \\
			\hspace*{5em}   -{}- {Equality does not always hold as choice removes dead code.}\\
			\hspace*{5em}   -{}- {Example $($equivalence, no equality$)$: \bl{p = $\langle$$\{[$1, 1$]\}$, $\pwrong \rangle$}.}
		\end{sdef}
	\end{bcolfr}
	\begin{bcolfr}[bot]
		\begin{sdef}{0.8}{choicerange}{Choice\_range}
			\hspace{2em}\bl{$\overline{\text{p} ~ \cup ~ \text{q}} ~ = ~ \overline{\text{p}}~  \cup ~ \overline{\text{q}}$}
			\\ \hspace*{5em}   -{}- {Choice distributes over the range of postconditions.}
		\end{sdef}
	\end{bcolfr}

	\begin{note}{Comment}
		These properties include the ``Laws of Choice'' of the Hoare theory papers (section \ref{hoare})\footnote{
			Except for the \bl{abort}-related property; see \ref{angelicdemonic} below.} --- but as (mechanically) proved theorems, not axioms.
	\end{note}


	\subsection{Properties of composition} \label{compositionproperties}

	\begin{bcolfr}[Theorems: Properties of composition]
		For programs \bl{p}, \bl{q}, \bl{v} and conditions \bl{C} and \bl{D}:\\

		\begin{sdef}{0.8}{composeassoc}{Compose\_assoc}
			\hspace{2em}\bl{p $;$ $($q $;$ v$)$ = $($p $;$ q$)$ $;$ v} \\
			\hspace*{5em}   -{}- {Composition is associative.}
			\\
			\hspace*{5em}   -{}- So we may apply ``\bl{$;$}'' to several operands \\
			\hspace*{5em}   -{}-  without parentheses, as in  \bl{p $;$ q $;$ v} \\
			\hspace*{5em}   -{}- and to a list of programs with  \bl{$\Sigma$}, see below.

		\end{sdef}
	\end{bcolfr}
	\begin{bcolfr}
		\begin{sdef}{0.7}{composeabsorbrest}{Compose\_absorbrest}
			\hspace{2em}\bl{C\textrm{:} $($p $;$ q$)$ = $($C\textrm{:} p$)$ $;$ q} \hspace*{5em}
			\\
			\hspace*{5em}   -{}- {Composition left-absorbs restriction.}
		\end{sdef}
	\end{bcolfr}
	\begin{bcolfr}
		\begin{sdef}{0.7}{composeabsorbcorest}{Compose\_absorbcorest}
			\hspace{2em}\bl{$($p $;$ q$)$ $\backslash$ D = p $;$ $($q $\backslash$ D$)$} \hspace*{5em}
			\\
			\hspace*{5em}   -{}- {Composition right-absorbs corestriction.}
		\end{sdef}
	\end{bcolfr}
	\begin{bcolfr}
		\begin{sdef}{0.8}{composechoiceleft}{Compose\_choiceleft}
			\hspace{2em}\bl{v $;$ $($p $\cup$ q$)$ $\equiv$ \bl{ $($v $;$ p$)$ $\cup$ $($v $;$ q$)$}}
			\\
			\hspace*{5em}   -{}- {Composition left-distributes over choice.}
			\\
			\hspace*{5em}   -{}- Equivalence only, not equality: choice removes dead code\\
			\hspace*{5em}   -{}- but composition may re-introduce some.\\
			\hspace*{5em}   -{}- Example: \bl{p = q = v = $\langle$$\{[$1, 1$], [$2, 1$]\}$, $\{1\}\rangle$}
		\end{sdef}
	\end{bcolfr}
	\begin{bcolfr}
		\begin{sdef}{0.7}{composechoiceright}{Compose\_choiceright}
			\hspace{2em}\bl{$($p $\cup$ q$)$ $;$ v $\equiv$ \bl{ $($p $;$ v$)$ $\cup$ $($q $;$ v$)$}}
			\\
			\hspace*{5em}   -{}- {Composition right-distributes right over choice}\\
			\hspace*{5em}   -{}- $($equivalence only$)$.
		\end{sdef}
	\end{bcolfr}
	\begin{bcolfr}[bot]
		\begin{sdef}{0.7}{composefeasibleleft}{Compose\_feasibleleft}
			\hspace{2em} If \bl{q} is feasible then \bl{p $;$ q} is feasible.
			\\
			\hspace*{5em}   -{}- {This result is stronger than \referencename{composefeasible}  $($\ref{presroundfeasible}$)$.}
		\end{sdef}
	\end{bcolfr}

	\begin{note}{Notation}%
		The composition \bl{p$_1$ $;$ ...; p$_n$} of a list of programs may be written \bl{$\Sigma$  p$_i$}. Associativity of composition $($\textcolor{green!50!black}{Compose\_assoc}$)$ justifies this notation. For an empty list, the sum is \bl{Skip} by convention $($see \ref{loopdef} for details$)$.
	\end{note}


	\subsection{Properties of roundness and feasibility} \label{presroundfeasible}
	\begin{colfr}[Theorems: Fundamental operators preserves roundness]

		For programs \bl{p} and \bl{q} and an arbitrary condition \bl{C}:\\
		\begin{sdef}{0.8}{roundrestrict}{Round\_restrict}
			\hspace{2em}\bl{$\lfloor$\bl{C}: \pr$\rfloor$ = $\lfloor$\pr$\rfloor$ / \bl{C}}
		\end{sdef}
		\begin{sdef}{0.8}{roundchoice}{Round\_choice}
			\hspace{2em}\bl{$\lfloor$p $\cup$ q$\rfloor$ = $\lfloor$p$\rfloor$ $\cup$ $\lfloor$q$\rfloor$}
		\end{sdef}
		\begin{sdef}{0.7}{roundcompose}{Round\_compose}
			\hspace{2em}\bl{$\lfloor$p $\textrm{;} $ q$\rfloor$ = $\lfloor$p$\rfloor$ \textrm{;} $\lfloor$q$\rfloor$}
		\end{sdef}
		\begin{sdef}{0.7}{roundcorestrict}{Round\_corestrict}
			\hspace{2em}\bl{$\lfloor$\pr$\backslash$\bl{C}$\rfloor$ = $\lfloor$\pr$\rfloor$ $\backslash$ \bl{C}}
		\end{sdef}
		\\
		As a consequence, if  \bl{p} and \bl{q} are rounded: \\
		\begin{sdef}{0.7}{restrictrounded}{Restrict\_rounded}
			\hspace{2em}\bl{C$:$} \bl{p} is rounded.
		\end{sdef}\\
		\begin{sdef}{0.8}{choicerounded}{Choice\_rounded}
			\hspace{2em}\bl{p $\cup$ q} is rounded. -{}- True even if  \bl{p} and \bl{q} are not rounded.
		\end{sdef}\\
		\begin{sdef}{0.7}{composerounded}{Compose\_rounded}
			\hspace{2em}\bl{p $;$  q} is rounded.
		\end{sdef}
	\end{colfr}


	\begin{note}{Proof idea}%
		For restriction $($\referencename{restrictrounded}$)$, the property to prove, assuming \bl{\underline{post$_p$} $\subseteq$ \pdomain{p}}, is \bl{\underline{post$_p$ / C} $\subseteq$ \pdomain{p} $\cap$ C}. It follows from the property of restriction on relations called \referencename{domainrestrict} in section \ref{relations}: \underline{\bl{r / X}} =  \underline{\bl{r}} $\cap$ \bl{X}.  For choice, if  \bl{\underline{post$_p$} $\subseteq$ \pdomain{p}} and \bl{\underline{post$_q$} $\subseteq$ \pdomain{q}} then \bl{\underline{post$_p$} $\cup$ \underline{post$_q$} $\subseteq$ \pdomain{p} $\cup$ \pdomain{q}};  rounding the operands of the union $($per the definition of the postcondition of the choice operation$)$ changes nothing since these operands are already rounded. Similarly, for composition, the postcondition is \bl{\post[p] $;$ \post[q]} and the theorem in this case follows from the property of composition of relations that \bl{\underline{r $;$ s} = \underline{r} $\cap$ r$^{-1}$ $($\underline{s}$)$}.
	\end{note}



	\begin{colfr}[Theorems: Fundamental operators preserve feasibility]
		For feasible \bl{p} and \bl{p$_2$} and an arbitrary condition \bl{C}: \\
		\begin{sdef}{0.7}{restrictfeasible}{Restrict\_feasible}
			\hspace{2em}\bl{C$:$} \bl{p} is feasible.
		\end{sdef}
		\vspace{0.05cm}\begin{sdef}{0.8}{choicefeasible}{Choice\_feasible}
			\hspace{2em}\bl{p $\cup$ q} is feasible.
		\end{sdef}
		\vspace{0.05cm}\begin{sdef}{0.7}{composefeasible}{Compose\_feasible}
			\hspace{2em}\bl{p} \bl{$;$} \bl{q} is feasible.
		\end{sdef}
	\end{colfr}

	\begin{note}{Proof idea}
		$($From definitions in \ref{basicops}$)$: for restriction, the precondition gets narrowed down $($strengthened$)$, preserving the feasibility property  \bl{\pdomain{p} $\subseteq$ \underline{post$_p$}}. For choice, the new precondition is the union of the preconditions, and the new postcondition is the union of the rounded postconditions; we apply \referencename{feasibleround} from \ref{trim}.
	\end{note}


	\subsection{A note about choice} \label{angelicdemonic}
	The choice operator is ``angelic'' (as expressed by the theorems \referencename{failchoice} and \referencename{failchoiceonly} in \ref{specialprops}) in the  sense that it will yield a result if either operand could do so.

	This behavior is what we usually want: if we are writing an Internet routing program which can  send a packet through either of two routers \bl{A} and \bl{B}, we will succeed if \textit{at least one} of them is up and running. The definition of choice (last column in the table defining basic operators in \referencename{choicedef}, section \ref{basicops})  reflects this property by using \bl{\pdomain{p} $\cup$ \pdomain{q} } as the precondition \bl{\pdomain{p $\cup$ q} } for the choice.

	Another variant is ``demonic'' choice, which  only guarantees a solution if \textit{every} operand can do so. If we are using someone else's packet delivery mechanism (rather than writing our own), and know that it may use either \bl{A} or \bl{B}, we can only assume that the packet will go through if \textit{both} servers are up.

	Demonic choice is easy to add to the present theory as a variant of choice in which the precondition of the choice is changed to \bl{\pdomain{p} $\cap$ \pdomain{q} }: the intersection, rather than the union, of the preconditions of the operands. A more complete exposition of the theory will include this variant.

	With its ``internal'' and ``external'' choice operators, the CSP concurrency calculus \cite{roscoecspnew} provides a distinction similar to angelic versus demonic. The explanation is that internal choice is made by the system and external by ``the environment''. These concepts are informal and, in practice, difficult to teach. A mathematical distinction --- union of preconditions versus their intersection --- seems preferable as it removes any uncertainty.

	The last axiom of the ``Laws of Choice'' by Hoare and colleagues cited earlier (\ref{hoare}) suggests that those authors' understanding of choice is demonic. This example provides further illustration of the dangers of a purely axiomatic approach: introducing both variants axiomatically would require a repetition of all the other choice axioms, since the operators share most of their other properties including associativity, commutativity and relationship with ``\bl{$;$}'' and other operators). With the axiom-free and theorem-oriented  approach of the present work, it suffices to add one definition for a new operator (specifying the new precondition); then its --- possibly numerous --- properties, instead of being tediously postulated, can be proved as theorems.

	\section{Refinement and related notions} \label{refinementgeneral}

	Programs and specifications are fundamentally the same, but the reason they have traditionally been distinguished from each other is that they may exist at widely varying levels of abstraction. If we consider programs and specifications \textit{relatively} to each other, the distinction makes sense: a program/specification may be a more concrete or more abstract version of another. In that case we might talk of the more concrete version as  the program and of the more abstract one as its specification. The notions of refinement, specialization and implementation formalize these observations.

	\subsection{Refinement: preliminary notions} \label{refinedefinitions}

	The following notions help define refinement and specialization.

	\begin{bcolfr}[Definition: state-extending, pre-weakening, post-strengthening] \label{strength}
		For programs \bl{q} and \bl{p}:\\
		\vspace{-0.1cm}
		\begin{sdef}{0.8}{extendstate}{Extend\_state}
			\hspace{2em}\bl{q} \bf{state-extends} \bl{p} if \bl{S$_p$ $\subseteq$ S$_q$}
		\end{sdef}
	\end{bcolfr}
	\begin{bcolfr}
		\begin{sdef}{0.8}{weakenpre}{Weaken\_pre}
			\hspace{2em}  \bl{q} \bf{pre-weakens} \bl{p} if \bl{\pdomain{p} $\subseteq$ \pdomain{q}}
		\end{sdef}
	\end{bcolfr}
	\begin{bcolfr}[bot]
		\begin{sdef}{0.8}{strengthenpost}{Strengthen\_post}
			\hspace{2em} \bl{q} \bf{post-strengthens} \bl{p} if \bl{$($\pdomain{p}$:$ \pdomain{q}$:$ post$_q$$)$ $\subseteq$ \post$_p$}
		\end{sdef}
	\end{bcolfr}

	\begin{note}{Caveat}%
		The terms ``extends'', ``weakens'' and ``strenghtens'' are convenient but should not mislead since all relationships use \bl{$\subseteq$}, not \bl{$\subset$}. So these terms should be understood as shorthand for ``extends or retains'' etc.
	\end{note}

	\begin{note}{Comment}%
		Post-strengthening $($strengthening the precondition$)$ means replacing the postcondition by a tighter one $($in the sense of \bl{$\subseteq$}, meaning one with as many or fewer input-output state pairs. This relation only applies, however, to input states that satisfy both preconditions. (A similar property applies to contract adaptation in redeclarations under inheritance in object-oriented programming \cite{oosc1}.)
	\end{note}

	Unlike with earlier discussions, the definitions make the state explicit, as refinement and its variants may introduce new $($concrete$)$ states in addition to the original's abstract states.

	\subsection{Refinement, specialization, implementation} \label{refinedef}

	With the preceding auxiliary  notions we can define three fundamental relations between programs:

	\begin{table}[H]
		\rowcolors{1}{bgcolor}{bgcolor}
		\arrayrulecolor{bordercolor}
		\setlength{\arrayrulewidth}{\borderwidth}
		\setlength{\tabcolsep}{12pt}
		\renewcommand{\arraystretch}{1.8}
		\begin{tabular}{|c|c|l|l l r|}
			\hline
			\bf{Notation}           & \bf{Name}                      & \shortstack{\bf{Alternative}                                                                               \\\bf{names}}       & \multicolumn{3}{l|}{\bf{Definition}}                                                           \\

			\hline
			\bl{q $\refines$ p}     & \bl{q} \bf{refines} \bl{p}     & \shortstack{\bl{p} \bf{specifies} $($or                                                                    \\ \bf{abstracts}$)$ \bl{q} } &\multicolumn{2}{l}{\shortstack[l]{\bl{q} post-strengthens \bl{p} \\ and \bl{q} pre-weakens \bl{p} \\ and \bl{q} state-extends \bl{p}}} & \hspace{-5pt} \defnraw{refinedef}{Refine\_def}\\
			\hline
			\bl{q $\specializes$ p} & \bl{q} \bf{specializes} \bl{p} & \bl{p} \bf{generalizes} \bl{q}          & \multicolumn{2}{l}{\shortstack[l]{\bl{q} post-strengthens \bl{p} \\ and \bl{p} pre-weakens \bl{q} \\ and \bl{p} state-extends \bl{q}}} & \hspace{-5pt}\defnraw{specialdef}{Special\_def}\\

			\hline
			                        & \bl{q} \bf{implements} \bl{p}  &                                         & \shortstack[l]{\bl{q}  is feasible                               \\ and refines  \bl{p}} & \multicolumn{2}{r|}{\defnraw{implementdef}{Implement\_def}}\\

			\hline
		\end{tabular}
		\label{tab:refine_special_impl}
	\end{table}

	\begin{note}{Comment}
		Because they are based on subsetting, the notations for refinement and specialization use variants of the subset symbol ``\bl{$\subseteq$}''. It is always clear from the operands whether ``\bl{$\subseteq$}'' denotes ``subset'' on sets or ``refine'' on programs.
	\end{note}

	\begin{note}{Justification}
		The difference between refinement and specialization is subtle but significant. Both strengthen the postcondition. Refinement --- like the rules governing inheritance in object-oriented programming --- accepts more states and a larger $($``weaker''$)$ precondition, meaning that the refined version $($\bl{q} where \bl{q $\refines$ p}$)$ may have new behavior for states not in covered by the original precondition. With specialization, in contrast, the new version $($\bl{q} where \bl{q $\specializes$ p}$)$ has fewer behaviors. The literature contains many studies of refinement, but some of them --- including by Hoare and colleagues \cite{hoare2012, hoare1997} --- cover what is here called specialization. Refinements also figures in formal specification approaches such as Z and B; its definition for Z in the main reference on the topic takes up 98 book pages (pages 53 to 150 of \cite{derrick2014refinement}), making comparisons hazardous. Refinement as defined here is in line with other approaches, particularly from Dijkstra and Wirth  \cite{dijkstra1968,dijkstra1972,dijkstra1976,wirth1971,wirth1973}.
	\end{note}

	\begin{note}{Notation caveat}
		The literature often uses \bl{q $\refines$ p} to express that \bl{p} refines \bl{q} rather than the converse as here. The convention  the present work $($where  \bl{q $\refines$ p} means that \bl{q} refines \bl{p}, and similarly for specialization$)$ is consistent with the defining property: for refinement as well as for specialization, the postcondition of \bl{q} is a subset of the postcondition of \bl{p}, not the other way around. $($A similarly unfortunate notational  paradox exists in logic, where ``\bl{p} implies \bl{q}'' is sometimes written \bl{p $\supset$ q} even though the implication means, if we view propositions as sets,  that \bl{p} is a subset of \bl{q} or equal.$)$
	\end{note}

	\subsection{Properties of refinement, specialization and implementation} \label{refineproperties}

	The relations just introduced --- refinement, specialization, implementation --- satisfy  some important properties.

	\begin{colfr}[Implementation theorem]
		\begin{sdef}{0.75}{implementfeasible}{Implement\_feasible}
			A program having an implementation is feasible.
		\end{sdef}
	\end{colfr}

	\begin{note}{Notation caveat}
		This important theorem states $($from the preceding definitions$)$ that if \bl{q} refines \bl{p} and is feasible, then \bl{p} itself is feasible.
	\end{note}

	\begin{note}{Comment}
		The  Implementation theorem actually works both ways: it can actually be extended to ``a program is feasible if and only if it has an implementation'', since the ``only if'' part is trivial $($if a program is feasible, then because of the reflexivity of refinement, seen below, it has an implementation --- itself$)$.
	\end{note}

	\begin{note}{Terminology}
		\begin{itemize}
			\item ``Order relation'' as used below denotes a non-strict possibly partial order: a relation ``\bl{$\leq$}'' that is reflexive $($\bl{x $\leq$ x} for all \bl{x}$)$, antisymmetric $($whenever \bl{x $\leq$ y} and \bl{y $\leq$ x} then \bl{x = y}$)$ and transitive $($whenever \bl{x $\leq$ y} and \bl{y $\leq$ z} then \bl{x $\leq$ z}$)$.
			\item
			      The relations under study --- ``refines'', ``specializes'', ``implements'' --- are antisymmetric only under equivalence. In other words, taking refinement as an example, if  \bl{p $\refines$ q} and \bl{q $\refines$ p} both hold, it is not necessarily the case that \bl{p = q} $($because of possible spurious elements, ``dead code'', on either side$)$, but equivalence \bl{p $\equiv$ q} does hold. If a relation has this property and is otherwise reflexive and transitive, we will say that, rather than \textit{being} an order relation, it \bf{induces} an order relation. $($Formally, the quotient of the relation by the equivalence relation ``\bl{$\equiv$}'' is an order relation.$)$
		\end{itemize}

	\end{note}

	\begin{colfr}[Theorems: order relations]
		\begin{sdef}{0.8}{refineorder}{Refine\_order}
			Refinement $($``\bl{$\refines$}''$)$ induces an order relation.\\
		\end{sdef}
		\begin{sdef}{0.8}{specialorder}{Special\_order}
			Specialization $($``\bl{$\specializes$}''$)$ induces an order relation\\
		\end{sdef}
		\begin{sdef}{0.7}{implementationorder}{Implementation\_order}
			Implementation induces an order relation\\
			$($except that it is reflexive for feasible elements only$)$
		\end{sdef}

	\end{colfr}

	\begin{note}{Comment}%
		For brevity, the separate components of each of the above properties $($reflexivity, antisymmetry, transitivity$)$ are not individually included. They figure in the Isabelle/HOL repository with their proofs, under self-explanatory names consistent with the general conventions, such as \defnfmt{Refine\_transitive}.
	\end{note}

	\begin{note}{Definition}
		The following theorems will express when refinement, specialization and implementation are compatible with basic operations. An operation will be called \textbf{refinement-safe} if applying it preserves refinement properties. Similarly: \bf{implementation-safe}, \bf{specialization-safe}.
	\end{note}

	\subsection{Refinement and basic operations} \label{refinesafe}

	As an example:

	\begin{colfr}[Theorem: Refinement safety $($for restriction$)$]
		\begin{sdef}{0.7}{restrictrefinesafety}{Restrict\_refinesafety}
			If \bl{q $\refines$ p}, then \bl{C$:$ q $\refines$ C$:$ p}\\
			\hspace*{4em} -{}- Restriction is refinement-safe.
		\end{sdef}
	\end{colfr}

	\begin{note}{Comment}
		On the other hand, many operations are not refinement-safe $($although many will be specialization-safe as seen below$)$ because of precondition widening. They include corestriction, composition and choice. In the last case a form of refinement-safety does hold per the following theorem.
	\end{note}

	\begin{colfr}[Theorem: Refinement safety for choice]
		\begin{sdef}{0.7}{choicerefinesafety}{Choice\_refinesafety}
			If \bl{q $\refines$ p} and \bl{q} post-strengthens \bl{v}, then \bl{q $\cup$ v} $\refines$ \bl{p $\cup$ v}
		\end{sdef}
	\end{colfr}

	\noindent Implementation safety follows directly from refinement safety.

	\subsection{Specialization and basic operations} \label{specialsafe}

	Unlike with refinement, all basic operations as well as corestriction are specialization-safe. \\

	\begin{bcolfr}[Theorems: Specialization safety]
		If \bl{q $\specializes$ p}, then the following properties hold:\\\\
		\begin{sdef}{0.7}{restrictspecialsafety}{Restrict\_specialsafety}
			\hspace*{2em} \bl{C$:$ q $\specializes$ C$:$ p}\\
			\hspace*{4em} -{}- Restriction is specialization-safe.
		\end{sdef}
	\end{bcolfr}
	\begin{bcolfr}
		\begin{sdef}{0.7}{corestrictspecialsafety}{Corestrict\_specialsafety}
			\hspace*{2em} \bl{C $\backslash$ q $\specializes$ C $\backslash$ p}\\
			\hspace*{4em} -{}- Coestriction is specialization-safe.
		\end{sdef}
	\end{bcolfr}

	\begin{bcolfr}
		\begin{sdef}{0.7}{choicerefsafety}{Choice\_specialsafety}
			\hspace*{2em} \bl{q $\cup$ v $\specializes$ p $\cup$ v}\\
			\hspace*{2em} \bl{v $\cup$ q $\specializes$ v $\cup$ p}\\
			\hspace*{4em} -{}- Choice is specialization-safe.
		\end{sdef}
	\end{bcolfr}

	\begin{bcolfr}[bot]
		\begin{sdef}{0.7}{composespecialsafety}{Compose\_specialsafety}
			\hspace*{2em} \bl{q $;$ v $\specializes$ p $;$ v}\\
			\hspace*{2em} \bl{v $;$ q $\specializes$ v $;$ p}\\
			\hspace*{4em} -{}- Composition is specialization-safe.
		\end{sdef}
	\end{bcolfr}


	\subsection{Restriction/corestriction under refinement and specialization} \label{refinerestrict}

	\begin{note}{Comment}
		Even though they are very close, having the same effect on the postcondition, refinement and specialization have distinct and sometimes inverse effects on restriction and specialization.
	\end{note}

	\begin{note}{Notation reminder}
		To avoid any confusion, note that ``\bl{$\subseteq$}'' in properties such as  \bl{D $\subseteq$ C} for subsets of the state set \bl{S} is the plain ``subset'' relation. On programs,``\bl{$\subseteq$}'' is specialization and ``\bl{$\subseteq$}'' is refinement. As another reminder, ``abstraction'' was defined in section $($\ref{refinedef}$)$ as  the inverse of refinement.
	\end{note}

	\begin{bcolfr}[Theorems: restriction and corestriction under refinement and specialization]
		For conditions \bl{C} and \bl{D} such that \bl{D $\subseteq$ C}:\\\\
		\begin{sdef}{0.7}{restrictspecial}{Restrict\_special}
			\hspace*{2em} \bl{C$:$ \pr $\specializes$ \pr}\\
			\hspace*{4em} -{}- Restriction is a form of specialization.\\

		\end{sdef}
	\end{bcolfr}

	\begin{bcolfr}
		\begin{sdef}{0.7}{restrictrefine}{Restrict\_refine}
			\hspace*{2em} \bl{\pr $\refines$ C$:$ \pr}\\
			\hspace*{4em} -{}- Restriction is a form of abstraction.\\
			\hspace*{4em} -{}- Note reversal of order from \referencename{restrictspecial}.
		\end{sdef}
	\end{bcolfr}

	\begin{bcolfr}
		\begin{sdef}{0.7}{corestrictspecial}{Corestrict\_special}
			\hspace*{2em} \bl{\pr$\backslash$ C $\specializes$ \pr}\\
			\hspace*{4em} -{}- Corestriction is a form of specialization.\\

		\end{sdef}
	\end{bcolfr}

	\begin{bcolfr}
		\begin{sdef}{0.7}{restrictrefineorder}{Restrict\_refineorder}
			\hspace*{2em}\bl{C$:$ \pr $\refines$ D: \pr}\\
			\hspace*{4em} -{}- Under refinement, restriction reverses subsetting.
		\end{sdef}
	\end{bcolfr}
	\begin{bcolfr}
		\begin{sdef}{0.7}{restrictspecialsubset}{Restrict\_specialsubset}
			\hspace*{2em} \bl{D: \pr $\specializes$ C$:$\ pr}\\
			\hspace*{4em} -{}- {Under specialization, restriction retains subsetting.}
		\end{sdef}
	\end{bcolfr}

	\begin{bcolfr} [bot]
		\begin{sdef}{0.7}{corestrictspecialsubset}{Corestrict\_specialsubset}
			\hspace*{2em} \bl{\pr $\backslash$ D $\specializes $ \pr $\backslash$ C}\\
			\hspace*{3em} -{}- {Under specialization, corestriction retains subsetting.}
		\end{sdef}
	\end{bcolfr}


	\begin{note}{Comment}
		The last two properties, governing specialization, have no direct counterparts  for refinement.
	\end{note}

	\subsection{Refinement/specialization and basic programs} \label{refinebasicprogs}

	\begin{bcolfr}[Theorems: Basic programs under refinement and specialization]
		\begin{sdef}{0.7}{refinespecial}{Refine\_special}
			\hspace*{2em} \bl{Infeasible $\refines$ Skip $\refines$ Havoc $\refines$ Fail}\\
			\hspace*{3em} -{}- {Full ordering of basic programs under refinement.}
		\end{sdef}
	\end{bcolfr}
	\begin{bcolfr}
		\begin{sdef}{0.7}{specialspecail}{Special\_special}
			\hspace*{2em} \bl{Fail $\specializes$ Infeasible $ \specializes $ Skip $ \specializes$ Havoc}\\
			\hspace*{3em} -{}- {Their $($different$)$ ordering under specialization.}

		\end{sdef}
	\end{bcolfr}

	\begin{bcolfr}
		\begin{sdef}{0.7}{refinehavoccompose}{Refine\_havoccompose}
			\hspace*{2em} \bl{\pr $\refines$ \pdomain{p}: Havoc}
		\end{sdef}
	\end{bcolfr}
	\begin{bcolfr}
		\begin{sdef}{0.7}{specialhavoccompose}{Special\_havoccompose}
			\hspace*{2em} \bl{\pr $\specializes$ \pdomain{p}: Havoc}
		\end{sdef}
	\end{bcolfr}
	\begin{bcolfr}
		\begin{sdef}{0.7}{refinehavoc2}{Refine\_havoc*}
			\hspace*{2em} \bl{\pr $\refines$ Havoc} if \pr\ is total.
		\end{sdef}
	\end{bcolfr}
	\begin{bcolfr}
		\begin{sdef}{0.7}{specialhavoc2}{Special\_havoc*}
			\hspace*{2em} \bl{\pr $\specializes$ Havoc}
		\end{sdef}
	\end{bcolfr}
	\begin{bcolfr}
		\begin{sdef}{0.7}{refinefail}{Refine\_fail}
			\hspace*{2em} \bl{\pr $\refines$ Fail}
		\end{sdef}
	\end{bcolfr}
	\begin{bcolfr}
		\begin{sdef}{0.7}{specialfail}{Special\_fail}
			\hspace*{2em} \bl{Fail $\specializes$ \pr}
		\end{sdef}
	\end{bcolfr}
	\begin{bcolfr}
		\begin{sdef}{0.7}{refinefailonly}{Refine\_failonly}
			\hspace*{2em} \bl{Fail $\specializes$ \pr} if and only if \pr $\equiv$ \bl{Fail}.
		\end{sdef}
	\end{bcolfr}
	\begin{bcolfr}[bot]
		\begin{sdef}{0.7}{specialfailonly}{Special\_failonly}
			\hspace*{2em}  \bl{\pr $\specializes$ Fail} if and only if \pr $\equiv$ \bl{Fail}.
		\end{sdef}
	\end{bcolfr}

	\begin{note}{Comment}
		Section \ref{basic} noted that the basic programs $($not only \bl{Skip}$)$ have variants such as  \bl{Fail$_C$} limited to specific state subsets. Here are the corresponding properties.
	\end{note}

	\begin{colfr}[Refinement and specialization properties of state-specific basic programs]
		For \bl{D $\subseteq$ C}:\\\\
		\begin{sdef}{0.7}{specialnonempty}{Special\_nonempty}
			\vspace{0.2cm}\hspace*{2em} \bl{Skip$_D \specializes$ Skip$_C$} \hspace*{5.6em} \bl{Skip$_C \refines$ Skip$_D$}\\
			\vspace{0.2cm}\hspace*{2em} \bl{Fail $_D \specializes$ Fail$_C$} \hspace*{5.5em} \bl{Fail $_C \refines$ Fail$_D$}\\
			\vspace{0.2cm}\hspace*{2em} \bl{Infeasible$_D \specializes$ Infeasible$_C$}\hspace*{1.5em} \bl{Infeasible$_C \refines$ Infeasible$_D$}\\
			\vspace{0.2cm}\hspace*{2em} \bl{Havoc$_D \specializes$ Havoc$_C$}\hspace*{4.2em} -{}- Not true for refinement.\\
			\vspace{0.2cm}\hspace*{2em} \bl{Fail$_{\pwrong}$ = Infeasible$_{\pwrong}$ = Skip$_{\pwrong}$ = Havoc$_{\pwrong} $}\\

		\end{sdef}
	\end{colfr}

	\begin{note}{Comment}
		These properties again illustrate the difference between refinement and specialization. Refinement may introduce new behavior by weakening the precondition.
	\end{note}


	\section{Conditional instructions} \label{conditionals}

	Among the basic operators, composition (``\bl{$;$}'') is directly reflected in programming languages by the sequencing operator, usually written with the same symbol. As noted in section \ref{basicops}, the other two, restriction and choice,  are fundamental building blocks but not generally used directly for programming. Actual programming languages generally use higher-level constructs: conditionals $($this section$)$ and loops $($section \ref{loops}).

	\subsection{Forms of conditional instruction} \label{conditionalforms}

	\begin{note}{Comment}
		A conditional instruction is a combination of choice and restriction. It applies to an arbitrary number of restricted programs, called \bf{branches}, with special case for one and two branches.
	\end{note}

	\begin{note}{Convention}
		In the following definitions and theorems:
		\begin{itemize}
			\item \bl{C}, \bl{D} and \bl{C$_1$, .. ,C$_n$} are conditions. \bl{C'} is the complement of \bl{C}.
			\item \bl{p}, \bl{q} and \bl{p$_1$,  .. , p$_n$} are programs.
		\end{itemize}
	\end{note}

	\begin{table}[H]
		\rowcolors{1}{bgcolor}{bgcolor}
		\arrayrulecolor{bordercolor}
		\setlength{\arrayrulewidth}{\borderwidth}
		\setlength{\tabcolsep}{12pt}
		\renewcommand{\arraystretch}{1.8}
		\begin{tabular}{|l| l r|}
			\hline
			\bf{Notation}                                          & \multicolumn{2}{l|}{\bf{Definition}}                                                            \\

			\hline
			\bl{\bf{if} C$_1$: p$_1$, ...,  C$_n$: p$_n$ \bf{end}} & \bll{$\bigcup_{~i:=1}^{~n}$ $C_i$: $p_i$}          & \defnraw{conditionalset}{Conditional\_set} \\

			\hline
			\bl{\bf{if} C \bf{then} p \bf{else} q \bf{end}}        & \bl{\bf{if} C$:$ p, C': q \bf{end}}                & \defnraw{conditionaltwo}{Conditional\_two} \\

			\hline
			\bl{\bf{if} C \bf{then} p \bf{end}}                    & \bl{\bf{if} C \bf{then} p \bf{else} Skip \bf{end}} & \defnraw{conditionalone}{Conditional\_one} \\

			\hline
		\end{tabular}
		\label{tab:conditional_definition}
	\end{table}

	\begin{note}{Definition}
		As noted, each \bl{ C$_i$: p$_i$} part of the basic conditional form is called a \bf{branch}.
	\end{note}

	\begin{note}{Extension}
		Not included, but easy to add, are variants with an arbitrary number of \bl{\bf{elseif}} clauses, with or without a final \bl{\bf{else}}.
	\end{note}

	\begin{note}{Caveat}
		All the variants given have the effect of a \bl{Skip} --- in other words, no effect --- if none of the conditions are satisfied. Dijkstra's version of the ``guarded form'' \bl{\bf{if} C$_1$: p$_1$, ...,  C$_n$: p$_n$ \bf{end}} fails in that case. This behavior can be obtained by introducing a variant produces \bl{Fail} if \bl{$\bigcup C_i$} has value \bl{False}. The versions with \bl{\bf{then}} are not affected since they cannot fail unless one of their branches does.
	\end{note}

	\subsection{Properties of conditional instructions} \label{conditionalprops}

	\begin{note}{Convention}
		For the purpose of the present discussion, the general conditional $($the first form above$)$ may be abbreviated \bl{\bf{if} C$_i$: p$_i$ \bf{end}}.
	\end{note}


	\begin{bcolfr}[Theorems: properties of conditionals]
		\begin{sdef}{0.7}{conditionalcommute}{Conditional\_commute}
			The conditional instruction \bl{\bf{if} C$_i$: p$_i$ \bf{end}} is equal to any other conditional obtained by permutation of its branches.
		\end{sdef}
	\end{bcolfr}
	\begin{bcolfr}
		\begin{sdef}{0.7}{conditionalsubspecial}{Conditional\_subspecial}
			If \bl{$D_i \subseteq C_i$} for all \bl{i} then
			\bl{$($\bf{if} D$_i$: p$_i$ \bf{end}$)$ $\specializes$ $($\bf{if} C$_i$: p$_i$ \bf{end}$)$}.
		\end{sdef}
	\end{bcolfr}
	\begin{bcolfr}
		\begin{sdef}{0.7}{conditionalmultsubspecial}{Conditional\_multsubspecial}
			If \bl{$q_i \specializes p_i$} for all \bl{i} then
			\bl{$($\bf{if} C$_i$: q$_i$ \bf{end}$)$ $\specializes$ $($\bf{if} C$_i$: p$_i$ \bf{end}$)$}.\\
			\hspace*{3em}   -{}- {No such properties for refinement }\\
			\hspace*{3em}   -{}- {$($same reason as non-safety of choice, see \ref{refinesafe}$)$.}

		\end{sdef}
	\end{bcolfr}
	\begin{bcolfr}
		\begin{sdef}{0.7}{conditionalsetone}{Conditional\_set\_one}
			\bl{C$:$ p} = \bl{{\bf{if} C$:$ p \bf{end}}}
			\\ \hspace*{3em}   -{}- {A plain \bl{\bf{if} ... \bf{then} ... \bf{end}} is just a restriction.}

		\end{sdef}
	\end{bcolfr}
	\begin{bcolfr}
		\begin{sdef}{0.7}{conditionalelements}{Conditional\_elements}
			\bl{C$_i$: p$_i$ $\specializes$ $($\bf{if} C$_1$: p$_1$, ...,  C$_n$: p$_n$ \bf{end}$)$} $($for \bl{$1 \leq i \leq n$}$)$.
			\\ \hspace*{3em}   -{}- {Every branch of a conditional specializes it.}
		\end{sdef}
	\end{bcolfr}

	\begin{bcolfr}[bot]

		\begin{sdef}{0.7}{conditionaldistrib}{Conditional\_distrib}
			\bl{D:  $($\bf{if} C$_i$: p$_i$ \bf{end}$)$~~  $\equiv$  ~~ \bf{if} $($D $\cap$ C$_i$$)$: p$_i$ \bf{end}}\\
			\hspace*{3em}   -{}- {Equivalence only because the order of applying}\\
			\hspace*{3em}   -{}- {restriction and choice matters for dead code.}
		\end{sdef}
	\end{bcolfr}


	\begin{note}{Justification}
		The following properties explain the use $($from \ref{conditions}$)$ of the alternative names ``\bl{True}'' for \bl{S} and \bl{False} for \bl{$\varnothing$}. Equivalence only.
	\end{note}

	\begin{bcolfr} [Theorems: properties of special conditions for conditionals]
		\begin{sdef}{0.8}{iftrue}{If\_true}
			\ifelse{True}{p$_1$}{p$_2$} $\equiv$ \pr[1]
		\end{sdef}
	\end{bcolfr}
	\begin{bcolfr} [bot]
		\begin{sdef}{0.8}{iffalse}{If\_false}
			\ifelse{False}{p$_1$}{p$_2$} $\equiv$ \pr[2]
		\end{sdef}
	\end{bcolfr}


	\section{Loops and invariants} \label{loops}

	\begin{note}{Justification}
		Two key capabilities of computers are: to execute a operation repeatedly $($and very fast$)$; and to evaluate a condition --- meaning formally, as we have seen, to perform a membership test between a given element $($a state$)$ and a given set. As a programming construct, the loop in its various forms captures this idea.
	\end{note}

	\subsection{Loops: definition} \label{loopdef}

	\begin{note}{Comment}
		To describe loops it is useful to start with a construct --- fixed repetition --- that reflects the standard mathematical practice of defining ``power'' as repeated self-application, and serves as a stepping stone for defining more general loops. $($In the following definitions, \bl{p} is an arbitrary program as usual.$)$
	\end{note}

	\begin{table}[H]
		\rowcolors{1}{bgcolor}{bgcolor}
		\arrayrulecolor{bordercolor}
		\setlength{\arrayrulewidth}{\borderwidth}
		\setlength{\tabcolsep}{6pt}
		\renewcommand{\arraystretch}{1.2}
		\begin{tabular}{|r|c|lr|}
			\hline
			\bf{Name}       & \bf{Notation}           & \multicolumn{2}{l|}{\bf{Definition}}                                                                    \\
			\hline
			Zero repetition & \bl{p$^0$}              & \bl{Skip$_{\underline{p}}$}                                        & \defnraw{zerorep}{Zero\_rep}       \\
			\hline
			\rule{0pt}{15pt}\rule{0pt}{40pt}\shortstack{Fixed                                                                                                   \\ repetition \\ $($or: ``power''$)$ \\ $($for \bl{n > 0$)$}}    & \shortstack{\bl{p$^n$} \\ $($Alternate notation: \\ \bl{\bf{for} i$:$ 1..n \bf{loop} p \bf{end}}}$)$                & \bl{$\Sigma_{~i = 1}^{~n}$ ~~ p} & \defnraw{fixedrep}{Fixed\_rep}         \\
			\hline
			\rule{0pt}{15pt}\rule{0pt}{25pt}\shortstack{Unbounded                                                                                               \\ repetition} & \aloop{p}                   & \bl{$\bigcup_{~i: ~ \mathbb{N}} ~~ p^i$}                                   & \defnraw{arbitraryrep}{Arbitrary\_rep} \\
			\hline
			``From'' loop   & \bl{\fromloop{a}{C}{b}} & \bll{\textit{a} $;$ $($\aloop{$\lnot $C$:$ p}$)$ $\backslash$ {C}} & \defnraw{fromloop}{From\_loop}     \\
			\hline
			``Repeat'' loop & \bl{\repeatloop{C}{b}}  & \bl{\fromloop{b}{C}{b}}                                            & \defnraw{whileloop}{While\_loop}   \\
			\hline
			``While'' loop  & \bl{\whileloop{C}{b}}   & \bl{\fromloop{Skip}{$\lnot$C}{b}}                                  & \defnraw{repeatloop}{Repeat\_loop} \\
			\hline
		\end{tabular}
		\label{tab:loop_operation}
	\end{table}

	\begin{note}{Justification}
		A loop repeats a certain program $($in practice, a program part$)$ subject to an exit condition, or equivalently a continuation condition $($these two forms of stating the condition are simply the negation of each other in the sense of the complement operation \bl{$``\lnot$}''$)$. The \bl{\bf{from}} and \bl{\bf{repeat}} forms loop until an exit condition holds; the \bl{\bf{while}} form loops until a continuation condition does not hold. Both \bl{\bf{from}} and \bl{\bf{while}} forms may iterate the body \bl{b} zero or more times; \bl{\bf{repeat}}, one or more times $($it starts by executing \bl{b} and only then asks whether to stop$)$. The \bl{\bf{from}} form includes an often useful initialization stage \bl{b}. All three forms are defined on the basis of the unbounded repetition construct $($\bl{\bf{loop} ... \bf{end}} which includes the zero-power \bl{p$^0$} among its possibilities. Although not common in programming languages, unbounded repetition exists for example in Ada.
	\end{note}

	\begin{note}{Comment}
		The  value for \bl{p$^0$} $($\referencename{zerorep}$)$ is required for consistency; see \referencename{powerinductive} below.
	\end{note}

	\begin{note}{Caveat}
		While an informal description of the fixed-repetition construct is that it ``executes \bl{p n} times'', in reality it executes it \textit{at most} \bl{n} times since the repetition will stop if after any number \bl{i} of iterations $($\bl{0 $\leq$ i $\leq$ n}$)$ \bl{\pdomain{p}} does not hold.
	\end{note}

	\begin{bcolfr}[Theorems: loop properties]
		\begin{sdef}{0.8}{powerinductive}{Power\_inductive}
			\begin{flalign*}
				 & \bll{p^{m + n} ~ = ~ p^{m} ~ ; ~ p^{n}} &
			\end{flalign*}
			\vspace{-.7cm}\\
			\hspace*{3em} -{}- For any \bl{m, n $\geq$ 0}.  For \bl{m = 0}, follows from\\
			\hspace*{3em} -{}- \referencename{zerorep}, \referencename{restrictown} and \referencename{skipcomprestrict}.
		\end{sdef}
	\end{bcolfr}
	\begin{bcolfr}
		\begin{sdef}{0.8}{loopchoice}{Loop\_choice}
			\begin{flalign*}
				 & \whileloop{C}{b} \equiv \bl{$\bigcup_{~i ~ \in ~ \mathbb{N}}$ $($C$:$ b$)^i$ ~ $\backslash$ ~$\lnot$ C} &
			\end{flalign*}
		\end{sdef}
	\end{bcolfr}
	\begin{bcolfr}
		\begin{sdef}{0.8}{repetitionfail}{Repetition\_fail}
			If \bl{$p^i \equiv$ Fail}, then \bl{$p^j \equiv$ Fail} for all \bl{$j \geq i$}\\
			\hspace*{3em} -{}- Program repetition cannot recover after any step has failed.
		\end{sdef}
	\end{bcolfr}
	\begin{bcolfr}
		\begin{sdef}{0.8}{loopfail}{Loop\_fail}
			\bl{$\bigcup_{1\leq i}^{n} p^i$ $\equiv$ Fail} implies \bl{$\bigcup_{1\leq i}^{m} p^i$ $\equiv$ Fail} for all \bl{m} and \bl{n}.
			\\ \hspace*{3em}{If a loop fails, it fails for any number of iterations.}
		\end{sdef}
	\end{bcolfr}
	\begin{bcolfr}
		\begin{sdef}{0.8}{looprange}{Loop\_range}
			\bll{\textoverline{\fromloop{a}{C}{b}} $\subseteq$ \bl{C}}\\
			\hspace*{3em} -{}- A terminating loop establishes its exit condition. See also \referencename{loopcorrect}.
		\end{sdef}
	\end{bcolfr}
	\begin{bcolfr}[bot]
		\begin{sdef}{0.8}{loopdistrib}{Loop\_distrib}
			If \bl{p} and \bl{q} are feasible, \bl{\textoverline{p} $\cap$ {\underline{q}} = $\varnothing$} and \bl{\textoverline{q} $\cap$ {\underline{p}} = $\varnothing$}, then:\\
			\hspace*{2em}\bll{(\aloop{p $\cup$ q}$)$ = (\aloop{p}$)$ $\cup$ (\aloop{q}}$)$\\
			\hspace*{3em} -{}- Under disjointness conditions, loop distributes over union.

		\end{sdef}
	\end{bcolfr}

	\begin{note}{Comment}
		The last result, \referencename{loopdistrib} can serve as a \textit{parallelization} law in advance of the discussion of concurrency (section \ref{concurrency}$)$.

	\end{note}


	\subsection{Invariants}

	\begin{note}{Notation reminder}
		\bl{r $($C$)$}, for a relation \bl{r}, is the image of a set \bl{C} under \bl{r}.
	\end{note}
	\begin{colfr}[Definition: invariants]
		\begin{sdef}{0.8}{programinv}{Program\_inv}
			A condition \bl{I} is an \bf{invariant} of a program \bl{p} if \bl{\textoverline{\post[p] (I $\cap$ \pdomain{p}}$)$ $\subseteq$ I}.
		\end{sdef}
	\end{colfr}
	\begin{note}{Justification}%
		An invariant is called that way because if it holds before application of \pr it will hold afterwards. More precisely, for the initial condition we need not the whole of \bl{I} but just \bl{I $\cap$ \underline{\pr}}, since results of \pr only matter when \pr starts in its precondition. The following theorems follow directly from the definition.
	\end{note}
	\begin{bcolfr}[Invariant properties]
		\begin{sdef}{0.8}{equivinv}{Equiv\_inv}
			If \bl{I} is an invariant of \bl{p} and \bl{\bl{p} $\equiv$ \bl{q}}, then \bl{I} is an invariant of \bl{q}.
		\end{sdef}
	\end{bcolfr}
	\begin{bcolfr}
		\begin{sdef}{0.8}{invdisjoint}{Inv\_disjoint}
			Any \bl{I} disjoint from \bl{\pdomain{p}} is an invariant of \pr.\\
		\end{sdef}
	\end{bcolfr}
	\begin{bcolfr}
		\begin{sdef}{0.8}{invtruefalse}{Inv\_truefalse}
			\bl{True} and \bl{False} are both invariants of \bl{p}.\\
		\end{sdef}
	\end{bcolfr}
	\begin{bcolfr}
		\begin{sdef}{0.8}{invsubset}{Inv\_subset}
			\bl{\textoverline{post$_p$}} and \bl{\prange{p}} are both invariants of \bl{p}.\\
			\hspace*{2em} -{}- Follow from properties of relations, particularly \referencename{imagerestrict}.
		\end{sdef}
	\end{bcolfr}
	\begin{bcolfr}
		\begin{sdef}{0.8}{invinter}{Inv\_inter}
			If \bl{I} and \bl{J} are invariants of \pr, then so are \bl{I $\cup$ J} and \bl{I $\cap$ J}.\\
			\hspace*{2em} -{}- Follow from \referencename{imageunion} and \referencename{imageinter}.

		\end{sdef}
	\end{bcolfr}
	\begin{bcolfr}
		\begin{sdef}{0.8}{invprop2}{Inv\_special}
			If \bl{I} is an invariant of \bl{p} and \bl{\bl{q} $\specializes$ p}, then \bl{I} is an invariant of \bl{\pdomain{p}$:$ \bl{q}}.
		\end{sdef}
	\end{bcolfr}
	\begin{bcolfr}[bot]
		\begin{sdef}{0.8}{invprop3}{Inv\_refines}
			If \bl{I} is an invariant of \bl{p} and \bl{\bl{q} $\refines$ \bl{p}}, then \bl{I} is an invariant of \bl{\bl{q}}.
		\end{sdef}
	\end{bcolfr}

	\noindent Some of the above results are cases of invariant-preserving operators in the following sense:\\

	\begin{bcolfr}[Definition and theorem: invariant-preserving operator, general invariant theorem]
		\begin{sdef}{0.8}{invpreserve}{Inv\_preserve}
			An operator on programs is \bf{invariant-preserving} if any invariant of all its operands is also an invariant of the operator's result.
		\end{sdef}
	\end{bcolfr}
	\begin{bcolfr}[bot]
		\begin{sdef}{0.8}{geninv}{Gen\_inv}
			All the program operators defined so far are invariant-preserving.
		\end{sdef}
	\end{bcolfr}

	\begin{note}{Comment}
		The operators covered are all those of section \ref{basicops}.
	\end{note}

	\vspace{-0.5cm}
	\subsection{Loop invariants} \label{loopinv}

	The general notion of invariant just introduced is particularly important in the case of loops.

	\begin{colfr}[Definition: loop invariant]
		\begin{sdef}{0.8}{loopinv}{Loop\_inv}
			A \bf{loop invariant} of \fromloop{a}{C}{b} is a subset of \bl{\textoverline{post$_a$}} that is an invariant of \bl{$\lnot$ C$:$ b}.
		\end{sdef}
	\end{colfr}

	\begin{note}{Justification}
		A loop invariant is ensured by the loop initialization (\bl{a}) and preserved by the loop body (\bl{b}$)$, and hence will hold after any execution of the loop as a whole.
	\end{note}

	\begin{note}{Notation reminder}
		\prange{p} is short for \bl{\textoverline{post$_p$ / \pdomain{p}}}: the set of states that \bl{p} actually produces (\ref{domainrangeprogram}$)$.
	\end{note}

	\begin{colfr}[Theorem: loop invariant properties]
		\begin{sdef}{0.8}{loopcorrect}{Loop\_correct}
			If \bl{I} is a loop invariant of the loop \bl{L = \fromloop{a}{C}{b}}, then \\
			\hspace*{3em}\bl{\prange{L}} \bl{$\subseteq$ C $\cap$ I}.
		\end{sdef}
	\end{colfr}

	\begin{note}{Justification}
		This theorem is the fundamental property of loop invariants\cite{meyer1980,furia2010}: the goal of a loop is to obtain on exit (\bl{\textoverline{L}}$)$ a combination of the exit condition (\bl{C}$)$ and a judiciously chosen invariant (\bl{I}, a weakening of the desired result$)$. From this observation follows a program construction method which splits the desired final condition into these two parts and produces the corresponding loop.
	\end{note}

	\begin{colfr}
		\begin{sdef}{0.8}{loopinvinv1}{Loop\_invinv}
			A loop invariant of \bl{L = \fromloop{a}{C}{b}} is also an invariant of \\
			\bl{L}, of  \bl{a}, and of  \bl{\aloop{$\lnot$ C$:$ b}}.
		\end{sdef}
	\end{colfr}

	\begin{note}{Comment}
		This theorem (which follows from \referencename{invsubset} above$)$ relates the notion of loop invariant to the more general notion of invariant.
	\end{note}



	\section{Contracted programs} \label{contracts}

	As noted at the beginning of section \ref{refinementgeneral}, the traditional distinction between specification and implementation is vague;  the formal definitions of refinement, implementation and specialization provided the first steps towards establishing it on solid ground. To obtain a complete picture, we must also consider the criterion of \textit {correctness}.

	It makes no sense to ask whether a program/specification, by itself is correct. Correct with respect to what? In the traditional world of programming, we do know what correctness means for a program: it performs according to a stated specification. A relative notion.

	What truly distinguishes a program from a specification, in the common usage of these terms, is neither just the level of abstraction nor the possibility of execution, but the existence of \bf{two} programs/specifications in the sense of the present theory, such that one of them is a refinement of the other. More precisely a \textit{feasible} refinement, or ``implementation'' (\ref{refinedef} and \referencename{implementfeasible}$)$. The following notation reflects this analysis.

	\begin{colfr}[Definition and notation: correctness, contracted program]
		\begin{sdef}{0.8}{correctdef}{Correct\_def}
			A program \bl{q} is \textbf{correct for} a program \bl{p} if it is an implementation of \bl{p}. The following \bf{contracted program} notation expresses this property:
			\par\setlength{\leftskip}{2em} \bl{\bf{require} \pdomain{p} \bf{do} q \bf{ensure} post$_p$ \bf{end}}
		\end{sdef}
	\end{colfr}

	\begin{note}{Comment}
		The property is transitive in the sense of the following theorem.
	\end{note}

	\begin{colfr}[Theorem: rule of consequence]
		\begin{sdef}{0.75}{ruleconsequence}{Rule\_consequence}
			A program \bl{v}, correct for \bl{q} which is correct for \bl{p}, is correct for \bl{p}.
		\end{sdef}
	\end{colfr}
	\begin{note}{Comment}
		The following concepts provide a different perspective on the preceding definitions. They refer to a program \bl{p}, a condition \bl{C} and a relation \bl{r} in \bl{S $\leftrightarrow$ S}.
	\end{note}

	\begin{bcolfr}[Definition: strongest postcondition, weakest precondition, ]
		\begin{sdef}{0.75}{strongespostcondition}{Strongest\_postcondition}
			\bl{\post[p] / C} is the \bf{strongest precondition} of \pr for \bl{C}, also written\\
			\hspace*{8.4em}\bl{\pr \bf{ sp} \textit{C}}
		\end{sdef}
	\end{bcolfr}
	\begin{bcolfr}[bot]
		\begin{sdef}{0.75}{weakestprecondition}{Weakest\_precondition}
			\bl{\pdomain{\pr} $-$ (\underline{\post[p] $-$ r}$)$} is the \bf{weakest precondition} of \pr for \bl{r}, also written\\
			\hspace*{8em} \bl{\pr \bf{ wp} r}
		\end{sdef}
	\end{bcolfr}
	\begin{note}{Explanation}
		\bl{\post[p] $-$ r} is the set difference of two relations, giving us the set of pairs that belong to the first but not to the second. Its domain, \bl{\underline{\post[p] $-$ r}}, is the set of states for which \pr produces at least one result that \bl{r} could never produce. Subtracting this domain from \bl{\underline{\pr}}, the precondition of \pr, gives us the full set of states satisfying the precondition on which \pr is guaranteed to agree with \bl{r}.
	\end{note}

	\begin{note}{Convention}
		In the following properties, \bl{p} and \bl{q} are programs, \bl{C} and \bl{D} sets of states, \bl{r} and \bl{t} relations in \bl{S $\leftrightarrow$ S}. While an \bl{\bf{sp}} expression takes a set as its second operand, a  \bl{\bf{wp}} expression needs a relation in \bl{S $\leftrightarrow$ S}; \bl{True$_{\leftrightarrow}$} will denote the full relation (also written  \bl{S $\times$ S} and \bl{Univ $[$S$]$})  and \bl{False$_{\leftrightarrow}$} the empty relation.
	\end{note}

	\begin{bcolfr}[Theorems: Properties of strongest postcondition and weakest precondition]
		If \bl{q} implements \bl{p}, then
		\par\setlength{\leftskip}{2em}\begin{sdef2}{0.6}{0.3}{postcharac}{Post\_charac}
			\bl{q \bf{sp} \pdomain{p} $\subseteq$ \post[p]}
		\end{sdef2}
		\par\setlength{\leftskip}{0em}and
		\par\setlength{\leftskip}{2em}\begin{sdef2}{0.6}{0.3}{precharac}{Pre\_charac}
			\bl{\pdomain{p} $\subseteq$ q \bf{wp} \post[p]}
		\end{sdef2}
		\hspace*{3em}-{}-  {Also expressible as \bl{q \bf{sp} \pdomain{p} $\Rightarrow$ \post[p]} and \bl{\pdomain{p} $\Rightarrow$ q \bf{wp} \post[p]}}
	\end{bcolfr}
	\begin{bcolfr}
		\begin{sdef}{0.8}{spfalse}{Sp\_false}
			\bl{\pr \bf{sp} False = False}
		\end{sdef}
	\end{bcolfr}
	\begin{bcolfr}
		\begin{sdef}{0.8}{wpfalse}{Wp\_false}
			\bl{\pr \bf{wp} False$_{\leftrightarrow}$ = False} for feasible \pr
		\end{sdef}
	\end{bcolfr}
	\begin{bcolfr}
		\begin{sdef}{0.8}{wpinfeas}{Wp\_infeas}
			\bl{\pr \bf{wp} False$_{\leftrightarrow}$ = \underline{\pr} $-$ \underline{\post[p]}}
		\end{sdef}
	\end{bcolfr}
	\begin{bcolfr}
		\begin{sdef}{0.8}{wpfail}{Wp\_fail}
			\bl{Fail \bf{wp} r = False}
		\end{sdef}
	\end{bcolfr}
	\begin{bcolfr}
		\begin{sdef}{0.8}{spfail}{Sp\_fail}
			\bl{Fail \bf{sp} C = False}
		\end{sdef}
	\end{bcolfr}
	\begin{bcolfr}
		\begin{sdef}{0.8}{sptrue}{Sp\_true}
			\bl{p \bf{sp} True = post$_p$}
		\end{sdef}
	\end{bcolfr}
	\begin{bcolfr}
		\begin{sdef}{0.8}{Wptrue}{Wp\_true}
			\bl{p \bf{wp} True$_{\leftrightarrow}$ = \pdomain{p}}
		\end{sdef}
	\end{bcolfr}
	\begin{bcolfr}
		\begin{sdef}{0.8}{sphavoc}{Sp\_havoc}
			\bl{Havoc \bf{sp} C = post (Havoc$)$ / C}
		\end{sdef}
	\end{bcolfr}
	\begin{bcolfr}
		\begin{sdef}{0.8}{spdistrib}{Sp\_distrib}
			\bl{\bl{p} \bf{sp} $($\bl{C} $\cup$ D$)$ = $($\bl{p} \bf{sp} \bl{C}$)$ $\cup$ $($\bl{p} \bf{sp} D$)$}
		\end{sdef}
	\end{bcolfr}
	\begin{bcolfr}[bot]
		\begin{sdef}{0.8}{wpdistrib}{Wp\_distrib}
			\bl{\bl{p} \bf{wp} $($\bl{r} $\cup$ t$)$ = $($\bl{p} \bf{wp} \bl{r}$)$ $\cup$ $($\bl{p} \bf{wp} t$)$}
		\end{sdef}
	\end{bcolfr}


	\vspace{-.3cm}
	\section{Concurrency}
	\label{concurrency}

	\vspace{-.2cm}
	One of the most important but also most delicate mechanisms of programming is \bf{concurrent
		operation}\footnote{Also known as “parallel”. There are technical differences between the
		two terms, which do not matter here.}. The reason
	concurrency is delicate is that it involves arbitrary \textit{interleaving}. As concurrency textbooks usually explain
	at the outset (see e.g. \cite{benari}$)$, when dealing with concurrency we do not need to assume  that two events
	ever take place at the same time, or even ask whether that is actually possible: we only have to contend with not knowing which comes first.
	So stating that two elementary operations \bl{a} and \bl{b} execute concurrently simply means that the execution could be \bl{a $;$ b} or \bl{b $;$ a}. This observation suggests a simple concurrency operator, although as we will see it is not general enough.



	\subsection{Definition: Atomic concurrency} \label{atomic}

	\begin{table}[H]
		\rowcolors{1}{bgcolor}{bgcolor}
		\arrayrulecolor{bordercolor}
		\setlength{\arrayrulewidth}{\borderwidth}
		\setlength{\tabcolsep}{12pt}
		\renewcommand{\arraystretch}{1.2}
		\begin{tabular}{|r|c|lr|}
			\hline
			\bf{Name}          & \bf{Notation}       & \multicolumn{2}{l|}{\bf{Definition}}                                             \\
			\hline
			Atomic concurrency & \bl{p \paralllel q} & \bl{$($p $;$ q$$)$$ $\cup$ $($q $;$ p$$)$$} & \defnraw{atomicconc}{Atomic\_conc} \\
			\hline
		\end{tabular}
		\label{tab:atomic_concurrency0}
	\end{table}




	\begin{note}{Comment}Atomic concurrency is not associative (as can be seen by noting for example that \bl{p $|||$ $($q $|||$ r$)$} includes among its choices  \bl{p $;$ r $;$ q}, which is not among the choices for \bl{$($p $|||$ q $)$ $|||$ r}). It can be generalized into an associative operator taking a list of operands, omitted here since it is not needed for the definition of the more general concurrency operator coming next. The following properties are the most important ones.
	\end{note}

	\begin{bcolfr}[Theorems: Properties of atomic concurrency]
		\begin{sdef}{0.8}{atomiccommute}{Atomic\_commute}
			\hspace{2em}\bl{p $|||$ q} = \bl{q $|||$ p}
			\\ \hspace*{5em} -{}- Atomic concurrency is commutative.
		\end{sdef}
	\end{bcolfr}
	\begin{bcolfr}
		\begin{sdef}{0.8}{atomicrestrict}{Atomic\_restrict}
			\hspace{2em}\bl{C$:$ $($p $|||$ q$)$} = \bl{$($ C$:$ p$)$ $|||$ $($ C$:$ q$)$}
			\\ \hspace*{5em} -{}- {Restriction distributes over atomic concurrency.}
		\end{sdef}
	\end{bcolfr}
	\begin{bcolfr}
		\begin{sdef}{0.8}{atomiccorestrict}{Atomic\_corestrict}
			\hspace{2em}\bl{$($p $|||$ q$)$ $\backslash$ C} = \bl{$($p $\backslash$ C$)$ $|||$ $($q $\backslash$ C$)$}
			\\ \hspace*{5em} -{}- {Corestriction distributes over atomic concurrency.}
		\end{sdef}
	\end{bcolfr}
	\begin{bcolfr}[bot]
		\begin{sdef}{0.8}{atomicfail}{Atomic\_fail}
			\hspace{2em}\bl{p $|||$ Fail = p $|||$Fail  = Fail}
			\\ \hspace*{5em} -{}- {\bl{Fail} absorbs atomic concurrency, left and right.}
		\end{sdef}
	\end{bcolfr}


	\subsection{Introducing a level of granularity} \label{granularity}

	The ``\bll{\paralllel}'' operator gives a general idea of what concurrency is about but is not appropriate for the
	practical uses of concurrency. The problem is that the programs we want to run in parallel, say \bl{p}
	and \bl{q}, are generally themselves composite programs, and a practical notion of concurrency must specify at what level of granularity we allow their \textit{components} to run in parallel.

\begin{figure}[hbt!]
		\centering
		\includegraphics[width=0.23\linewidth]{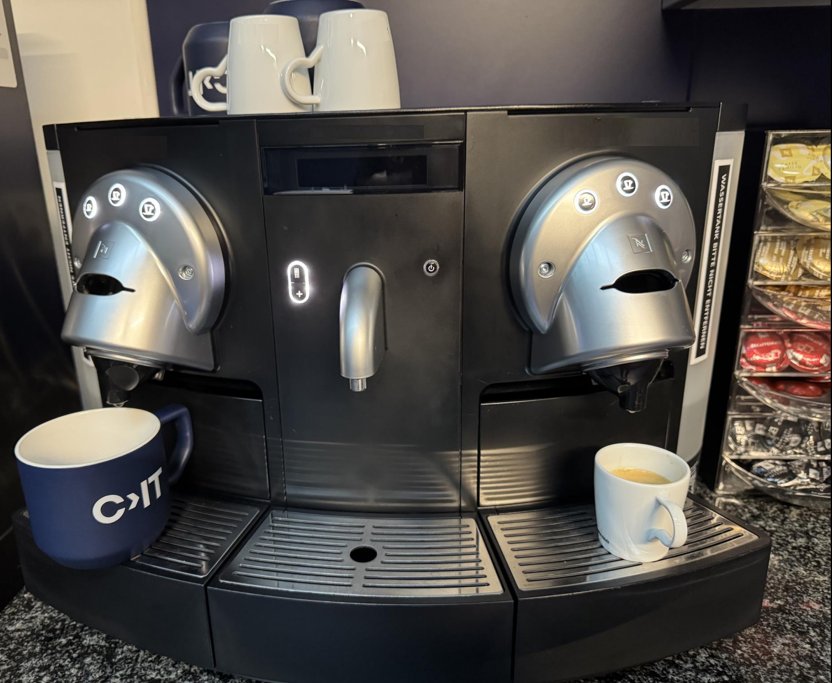}
		\caption{\bf{A parallel coffee machine}}\label{coffee}
	\end{figure}

	If \bl{p} = \bl{$($a $;$ b$)$} and \bl{q} = \bl{$($c $;$ d$)$}, then  \bl{p \paralllel q}
	only includes two executions: \bl{$($a $;$ b $;$ c $;$ d$)$} and \bl{$($c $;$ d; a $;$ b$)$}. True concurrent execution
	would also include all interleavings in which all four appear as long as \bl{a} is before \bl{b} and \bl{c} before
	\bl{d}, including \bl{$($c $;$ a $;$ b $;$ d$)$}, \bl{$($a $;$ c $;$ b $;$ d$)$} etc. As a simple example, consider programs using one of the most important resources of Constructor Institute of Technology, shown in Fig. \ref{coffee} and supporting a variant of Alfred Renyi's maxim\footnote{The original is about one mathematician and is often wrongly ascribed to Renyi's friend Paul Erdös.}: two software engineers are a device for turning coffee into a theory of programming.

	\vspace{-.3cm}
	\noindent The program \bl{p} could be

	\begin{itemize}
		\item []
		      \bl{put\_capsule\_left $;$
			      push\_button\_left $;$
			      get\_coffee\_left}
	\end{itemize}
	\vspace{-.1cm}
	and \bl{q} the same with \bl{left} replaced by \bl{right}. Then \bl{p \paralllel q} represents making two coffees one after the other, the only flexibility being afforded by allowing either order. In practice, however, the machine is ``more parallel'' than that: it allows any interleaving of the operations on both sides,  \bl{put\_capsule\_x}, \bl{push\_button\_x} and \bl{get\_coffee\_x} where \bl{x} is \bl{left} or \bl{right}, separately preserving the given order for each \bl{x}. $($Practitioners of concurrency know that such a finer level of granularity can lead to problems, known as ``data races'', if these operations use shared resources, but we are not there yet.$)$ As one of the variants tested as part of the preparation of this article, the following sequence, not covered by \bl{p \paralllel q}, was experimentally shown to work\footnote{Order is preserved between the successive composition components of each operand of the $\paralllel$, but not between those of different operands. The property, established for full concurrency as a theorem below \referencename{concsound}, is a general feature of concurrency: one cannot assume the existence of a global clock. In this particular execution trace, while the left button was pushed first, the corresponding coffee came out second. Perhaps left wanted a large coffee and right a small one.}:

\begin{itemize}
	\item []
	      \bl{
		      \bl{
			      \hspace*{0em}put\_capsule\_right $;$ put\_capsule\_left $;$ push\_button\_left $;$   \\
			      \hspace*{0.7em}push\_button\_right $;$  get\_coffee\_right $;$ get\_coffee\_left}
	      }
\end{itemize}




\noindent These observations indicate that obtaining a general notion of concurrency
requires choosing the appropriate level of granularity. We must accordingly express programs in a specific form.
\vspace{-.3cm}

\subsection {Civilized programs and Choice Normal Form} \label {CNF}

\begin{bcolfr}[Definitions: basic program, basis, civilized program]
	\begin{sdef}{0.8}{civilizeddef}{Civilized\_def}
		Given a set \bl{B} of ``\bf{basic programs}'' which includes \bl{Skip} and \bl{Fail}, a program is \bf{civilized} with \bf{basis} \bl{B} if it of one of the following forms, where \bl{p} and \bl{q} are $($recursively$)$ civilized:
	\end{sdef}
\end{bcolfr}
\begin{bcolfr}
	\begin{sdef}{0.8}{civilizedbasic}{Civilized\_basic}
		\begin{itemize}
			\item An element of \bl{B}
		\end{itemize}
	\end{sdef}
\end{bcolfr}
\begin{bcolfr}
	\begin{sdef}{0.8}{civilizedchoice}{Civilized\_choice}
		\begin{itemize}
			\item \bl{p $\cup$ q}
		\end{itemize}
	\end{sdef}
\end{bcolfr}
\begin{bcolfr}
	\begin{sdef}{0.8}{civilizedsequence}{Civilized\_sequence}
		\begin{itemize}
			\item \bl{p $;$ q}
		\end{itemize}
	\end{sdef}
\end{bcolfr}
\begin{bcolfr}[bot]
	\begin{sdef}{0.8}{civilizedrestrict}{Civilized\_restrict}
		\begin{itemize}
			\item \bl{ C$:$ p}
		\end{itemize}
	\end{sdef}
\end{bcolfr}

\noindent The rest of this discussion assumes that we are dealing with civilized programs, for some basis. This assumption entails no loss of generality, since we can always include all the mechanisms we want --- if not covered by sequence, choice or restriction --- in the basis. (But we will not then be able to interleave their components.)

Any civilized program can be expressed in the following \textit{normal form}:
\begin{colfr}[Definition and theorem: Choice Normal Form $($CNF$)$]
	\begin{sdef}{0.8}{normalform}{Normal\_form}
		Every civilized program can be expressed in Choice Normal Form as:
		\bll{$$\bigcup_{i: ~ I} \ ~~ \sum_{j:~ J_i} ~~p_i^j$$}
		where every \bl{p$_i^j$} is a basic program possibly with a restriction.
	\end{sdef}
\end{colfr}

\begin{note}{Explanation}
	``With a restriction'' means of the form \bl{ C$:$ p} for a condition \bl{C}.
	The \bl{J$_i$} index sets are finite;  \bl{I} can be infinite, but the discussion will limit itself to the finite case.
\end{note}

\begin{note}{Justification}
	Intuitively, the Choice Normal Form represents expressing a program as the set of all its possible execution paths.
\end{note}
\begin{note}{Terminology}
	Each \bl{$\Sigma$} term of the overall \bl{$\bigcup$} will be called a \bf{sum}. An empty sum denotes \bl{Fail}. For non-empty sum  \bl{s}, \bl{s$^1$} is the first element and \bl{s$^-$} the remaining (possibly empty) sum after removing the first element.
\end{note}

\subsection{Fine-grained concurrency} \label{concurrencyfine}
On civilized programs, we can define a meaningful concurrency operator ``\bl{$||$}'', which extends the atomic form ``\bl{$|||$}'' by taking into account a chosen level of granularity.

\begin{bcolfr}[Definition: Concurrent operation of two programs]
	The value of both \bl{p $||$ q} and \bl{q $||$ p} is defined as follows by structural induction on the definition of civilized programs (\referencename{civilizeddef}$)$, always applying the first rule that matches:
\end{bcolfr}
\begin{bcolfr}
	\begin{sdef}{0.8}{concbasic}{Conc\_basic}
		\begin{itemize}
			\item If both \bl{p} and \bl{q} are basic, possibly with a restriction, then \bl{p $|||$ q}.\\
			      \hspace*{1cm}-{}- This case uses atomic concurrency as defined in \ref{atomic}: \bl{$($p $;$ q$)$ $\cup$ $($q $;$ p$)$}.
		\end{itemize}
	\end{sdef}
\end{bcolfr}
\begin{bcolfr}
	\begin{sdef}{0.8}{concrestrict}{Conc\_restrict}
		\begin{itemize}
			\item If \bl{p} is \bl{\bl{C$:$ } x}, then \bl{$($p' $||$ q$)$} where \bl{p'} is \bl{p} with one restriction removed through \referencename{restrictinter}, \referencename{restrictdistrib} or \referencename{composeabsorbrest}.
		\end{itemize}
	\end{sdef}
\end{bcolfr}
\begin{bcolfr}
	\begin{sdef}{0.8}{concchoice}{Conc\_choice}
		\begin{itemize}
			\item
			      If \bl{p} is \bl{u $\cup$ v}, then \bl{$($p $||$ u$)$ $\cup$ $($p $||$ v$)$}.
		\end{itemize}

	\end{sdef}
\end{bcolfr}
\begin{bcolfr}[bot]
	\begin{sdef}{0.8}{concsum}{Conc\_sum}
		\begin{itemize}
			\item If both \bl{p} and \bl{q} are sums, then\\
			      \hspace*{3em}\bl{$($p$^1$ $;$ $($p$^-$ $||$ q$)$$)$ ~ $\cup$ ~$($q$^1$ $;$ $($p $||$ q$^-$$)$$)$}.
		\end{itemize}
	\end{sdef}
\end{bcolfr}

\begin{note}{Notation}
	A frequently encountered notation for \bl{p $||$ q $||$ ... } in languages supporting concurrency is \bl{\bf{parbegin} p, q, .... \bf{end}}, or some other keyword instead of \bl{\bf{parbegin}}. The number of operands can be arbitrary thanks to the associativity of the operator, \referencename{concassoc} below.
\end{note}

\begin{note}{Explanation}
	As noted, the cases have to be applied in order, using the first applicable one, the components then recursively subjected to the same rules. The four cases reflect the notion of interleaving:
	\begin{itemize}
		\item For basic operands \referencename{concbasic}, parallel execution means that either of the two possible orders is possible.
		\item Case \referencename{concbasic} can only apply to non-basic \bl{p} (otherwise the preceding case would capture it). Then \bl{p} must involve another operator; the rules on restriction will remove one restriction operator, enabling recursive application of the remaining cases (in the end restriction will only apply to basic operators, triggering the previous case).
		\item Executing \bl{p} in parallel with either \bl{u} or \bl{v} means either executing \bl{p} in parallel with \bl{u} or executing \bl{p} in parallel with \bl{v} --- as captured by \referencename{concchoice}.
		\item As a result of the repeated recursive application of the previous cases we are only left with the case \referencename{concsum} of parallel execution of two \bl{$\Sigma$} sums \bl{p} and \bl{q}. (If the operator on either side is any other than ``\bl{$;$}'', one of the other cases kicks in.) Then parallel execution means that we have a choice between only two possibilities: start with the first step of \bl{p} and continue with the rest of \bl{p} in parallel with all of \bl{q}; and the other way around. Each is of course subject to recursive expansion through the four cases.
	\end{itemize}
	Defining the size of an expression as the number of its operators, every case reduces the size of at least one operand of every ``\bl{$||$}'' subexpression, ensuring that the recursion will eventually terminate.
\end{note}

\begin{note}{Example}
	Consider \bl{$($u $\cup$ v$)$ $||$ $($x $;$ y$)$} (or this expression with its operands of the ``\bl{$\cup$}'' switched):
	\begin{itemize}
		\item It is subject to \referencename{concchoice}, not \referencename{concsum}. The rule yields \bl{$($u $||$ $($x $;$ y$)$$)$ $\cup$ $($v $||$ $($x $;$ y$)$$)$}.
		\item  Each of the two operands of the resulting choice is a parallel application of two sums (with one of the sums having only one element). Assuming all operands are basic, applying \referencename{concsum} twice yields
		      \bl{$($$($u $;$ x $;$ y$)$ $\cup$ $($x $;$ u $;$ y$)$ $\cup$ $($x $;$ y $;$ u$)$$)$ $\cup$ $($$($v $;$ x $;$ y$)$ $\cup$ $($x $;$ v $;$ y$)$ $\cup$ $($x $;$ y $;$ v$)$$)$}.
		\item Removing unneeded parentheses (per associativity of choice \referencename{choiceassoc}) puts this result in CNF.
	\end{itemize}
\end{note}

\begin{note}{Comment}
	The last example illustrates a fundamental property of the parallel operator expressed by the following theorem, which expresses when in parallel execution we can expect an original order of components to be preserved, and when not. The theorem relies on the notion of ``occurs before'' and ``interleaved''.
\end{note}
\begin{note}{Terminology}
	\begin{itemize}
		\item
		      A (civilized) program \bl{q} is ``\bf{part of}'' another if it is that program or (recursively) part of one of its operands in the corresponding case of the definition of civilized programs (\referencename{civilizeddef}).
		\item \bl{u} ``\bf{occurs before} \bl{v} in \bl{p}'' if  some part of \bl{p} is of the form \bl{u' $;$ v'} where \bl{u} is part of \bl{u'} and \bl{v} part of \bl{v'}.
		\item
		      \bl{p} and \bl{q} are ``\bf{interleaved}'' in \bl{v} if both \bl{p} occurs before \bl{q} and conversely in \bl{v}.
	\end{itemize}

	\noindent For a program in CNF, ``occurs before'' can be stated non-recursively: \bl{u} is an element of one of the sums of \bl{p}, and \bl{v} an element of that same sum, with a higher index. This definition, however, is more restricted since it only applies to basic programs.
\end{note}

\subsection{Concurrency properties} \label{concurrencyproperties}

\begin{colfr}[Theorem: Concurrency soundness]
	\begin{sdef}{0.8}{concsound}{Conc\_sound}
		If \bl{u} occurs before \bl{v} in \bl{p} and \bl{x} appears before \bl{y} in \bl{q}, then in \bl{p $||$ q}:
	\end{sdef}

	\vspace{.2cm}

	\begin{sdef}{0.7}{concsoundpreserve}{Conc\_soundpreserve}
		\begin{itemize}
			\item These two ``occurs before'' properties also hold.
		\end{itemize}
	\end{sdef}

	\vspace{.2cm}

	\begin{sdef}{0.7}{concsoundinterleave}{Conc\_soundinterleave}
		\begin{itemize}
			\item Each of \bl{u} and \bl{v} is interleaved with each of \bl{x} and \bl{y}.
		\end{itemize}
	\end{sdef}

\end{colfr}

\begin{note}{Proof}
	Structural induction on the definition of ``\bl{$||$}''. The theorem holds for basic programs (\referencename{concbasic}) and is preserved by each of the other cases if it holds for the operands.
\end{note}

\begin{note}{Comment}
	The theorem is also notable for what it does not guarantee: any order property between individual components of \bl{p} and \bl{q}. Concurrency does not assume any global clock. Order of execution is preserved between sequentially ordered parts of each operand of the concurrency (each ``concurrent process'' in frequently used terminology), but there is no guaranteed order between parts of different operands.
\end{note}

\noindent In addition to this fundamental theorem, concurrent operation satisfies a number of important properties:

\begin{bcolfr}[Theorems: concurrency properties]
	\begin{sdef}{0.8}{conccommute}{Conc\_commute}
		\hspace{2em}\bl{p $||$ q} = \bl{q $||$ p}
		\\ \hspace*{5em} -{}- {Concurrency is commutative.}
	\end{sdef}
\end{bcolfr}
\begin{bcolfr}
	\begin{sdef}{0.8}{concassoc}{Conc\_assoc}
		\hspace{2em}\bl{$($p $||$ q$)$ $||$ u = p $||$ $($q $||$ u$)$}
		\\ \hspace*{5em} -{}- {Concurrency is associative.}
	\end{sdef}
\end{bcolfr}
\begin{bcolfr}
	\begin{sdef}{0.8}{concchoicedistrib}{Conc\_choicedistrib}
		\hspace{2em}\bl{p $||$ $($q $\cup$ u$)$} = \bl{$($p $||$ q $)$ $\cup$ $($p $||$ u$)$}
		\\ \hspace*{5em} -{}- {Concurrency distributes over choice.}
	\end{sdef}
\end{bcolfr}
\begin{bcolfr}
	\begin{sdef}{0.8}{conccomposeleft}{Conc\_composeleft}
		\hspace{2em}\bl{$($p $||$ q$)$ $;$ u ~ $\subseteq$ ~ p $||$ $($q $;$ u$)$}
	\end{sdef}
\end{bcolfr}
\begin{bcolfr}
	\begin{sdef}{0.7}{conccomposeleftright}{Conc\_composeleftright}
		\hspace{2em}\bl{q $;$ $($p $||$ u$)$ ~ $\subseteq$ ~ p $||$ $($q $;$ u$)$}
	\end{sdef}
\end{bcolfr}
\begin{bcolfr}
	\begin{sdef}{0.8}{conccomposeright}{Conc\_composeright}
		\hspace{2em}\bl{p $;$ $($q $||$ u$)$ ~$\subseteq$ ~ $($p $;$ q$)$ $||$ u}
	\end{sdef}
\end{bcolfr}
\begin{bcolfr}
	\begin{sdef}{0.7}{conccomposerightleft}{Conc\_composerightleft}
		\hspace{2em}\bl{$($p $||$ u$)$ $;$ q ~ $\subseteq$ ~ $($p $;$ q$)$ $||$ u}
	\end{sdef}
\end{bcolfr}
\begin{bcolfr}[bot]
	\begin{sdef}{0.7}{conccomposegeneral}{Conc\_composegeneral}
		\hspace{2em}\bl{$($p $||$ q$)$ $;$ $($u $||$ v$)$ ~ $\subseteq$ ~ $($p $;$ u$)$ $||$ $($q $;$ v$)$}
	\end{sdef}
\end{bcolfr}

\begin{note}{Comment}
	These theorems --- in particular \referencename{conccomposegeneral} --- cover the ``axiom of exchange'' and all the other ``laws of exchange'' of the ``Laws of Programming'' work of Hoare and colleagues.

\end{note}

\section{Conclusion and further work}
\label{conclusion}


The PRISM theory outlined above illustrate the two principal theses of the present work: that programming does not need a complex mathematical apparatus (as in most formal methods work), but is entirely definable as the combination of one relation and one set, on the sole basis of elementary set-theory concepts; and that such a description requires no new axioms whatsoever, all relevant properties being provable (and actually proved through a mechanical prover as part of the production of this article).

The part of the theory that we have presented  covers fundamental programming constructs: the  classical control structures of ``structured programming'' (sequence, conditionals and loops) as well as a general framework for concurrency. Work still lies ahead to cover other programming mechanisms, including those of imperative programming languages (such as assignment), functional programming languages, object-oriented programming, as well as specialized mechanisms such as exception handling. The theory can also serve as the basis for numerous programming tools and for providing facilities for trying out, compiling, interpreting and verifying programming ideas and constructs.

\vspace{0.5cm}

\begin{note}{Acknowledgments}
	While section \ref{overview} has pinpointed perceived limitations of classic formal-methods approaches, particularly axiomatic ones, this work is obviously based on their insights and indebted to their pioneering authors. A talk on the topic to the IFIP WG2.3 working group elicited important suggestions by Mark Utting, Gary Leavens and other members.
\end{note}



\bibliographystyle{plain}
\bibliography{biblio}

\begin{thebibliography}{10}

\bibitem{benari}
Jean-Raymond Abrial.
\newblock {\em The B-book}.
\newblock Cambridge university press, 1996.

\bibitem{back2012}
Ralph-Johan Back and Joakim Wright.
\newblock {\em Refinement calculus: a systematic introduction}.
\newblock Springer Science \& Business Media, 2012.

\bibitem{abrial1996}
Mordechai Ben-Ari.
\newblock {\em Principles of Concurrent and Distributed Programming, 2nd edition}.
\newblock Addison-Wesley, 2005.

\bibitem{dijkstra1972}
Ole-Johan Dahl, Edsger~Wybe Dijkstra, and C.~A.~R. Hoare.
\newblock {\em Structured programming}.
\newblock Academic Press Ltd., 1972.

\bibitem{derrick2014refinement}
John Derrick and Eerke~A. Boiten.
\newblock {\em Refinement in Z and Object-Z: Foundations and Advanced Applications}.
\newblock Formal Approaches to Computing and Information Technology (FACIT). Springer, London, 2 edition, 2014.

\bibitem{dijkstra1976}
E.~W. Dijkstra.
\newblock {\em A discipline of programming}.
\newblock Prentice-Hall, Englewood Cliffs, NJ, 1976.

\bibitem{dijkstra1968}
Edsger~W Dijkstra.
\newblock A constructive approach to the problem of program correctness.
\newblock {\em BIT Numerical Mathematics}, 8(3):174--186, 1968.

\bibitem{dijkstra1974}
Edsger~W Dijkstra.
\newblock Ewd 463: Some questions, 1974.

\bibitem{dijkstra1975}
Edsger~W. Dijkstra.
\newblock Guarded commands, nondeterminacy and formal derivation of programs.
\newblock {\em Commun. ACM}, 18(8):453–457, August 1975.

\bibitem{floyd1967}
Robert~W. Floyd.
\newblock Assigning meanings to programs.
\newblock In {\em Proceedings of Symposium on Applied Mathematics}, volume~19, pages 19--32, 1967.

\bibitem{furia2010}
Carlo~Alberto Furia and Bertrand Meyer.
\newblock Inferring loop invariants using postconditions.
\newblock {\em Fields of Logic and Computation: Essays Dedicated to Yuri Gurevich on the Occasion of His 70th Birthday}, pages 277--300, 2010.

\bibitem{hehner}
Eric~C.R. Hehner.
\newblock {\em A Practical Theory of Programming}.
\newblock Springer, 1993.

\bibitem{hoare1997}
C.~A.~R. Hoare.
\newblock Unified theories of programming.
\newblock In {\em Mathematical methods in program development}, pages 313--367. Springer, 1997.

\bibitem{hoareslides}
C.~A.~R. Hoare.
\newblock Laws of programming with concurrency (slides).
\newblock In {\em MSR Concurrency Workshop}, 06 2013.

\bibitem{hoare1987}
C.~A.~R. Hoare, Ian Hayes, He~Jifeng, Carroll Morgan, A.~Roscoe, Jeff Sanders, Ib~Sørensen, J.~Spivey, and Bernard Sufrin.
\newblock Laws of programming.
\newblock {\em Communications of the ACM}, 30:672--686, 08 1987.

\bibitem{hoare2018}
C.~A.~R. Hoare, Alexandra Mendes, and João~F. Ferreira.
\newblock Algebra, logic, geometry: at the foundations of computer science.
\newblock In {\em Formal Methods Teaching, LNCS 11758}, pages 3--20. Springer, 2018.

\bibitem{hoarepraise}
C.~A.~R. Hoare and Stephan van Staden.
\newblock In praise of algebra.
\newblock {\em Formal Aspects of Computing}, 24:423--431, 2012.

\bibitem{hoare2012}
C.~A.~R. Hoare and Stephan van Staden.
\newblock The laws of programming unify process calculi.
\newblock In Jeremy Gibbons and Pablo Nogueira, editors, {\em Mathematics of Program Construction}, pages 7--22. Springer, 2012.

\bibitem{kahn1987}
Gilles Kahn.
\newblock Natural semantics.
\newblock In {\em Symposium on Theoretical Aspects of Computer Science}, 1987.

\bibitem{Shaoying}
Shaoying Liu.
\newblock Extending operation semantics to enhance the applicability of formal refinement.
\newblock In {\em Specification, Algebra and Software}, volume LNCS 8373, pages 434--440. Springer, 2014.

\bibitem{mendelson}
Eliot Mendelson.
\newblock {\em Introduction to Mathematical Logic (6th edition)}.
\newblock Chapman and Hall, 1976.

\bibitem{meyer1980}
Bertrand Meyer.
\newblock A basis for the constructive approach to programming.
\newblock In {\em IFIP Congress}, pages 293--298, 1980.

\bibitem{oosc1}
Bertrand Meyer.
\newblock {\em Object-Oriented Software Construction (first edition)}.
\newblock Prentice Hall, 1988.

\bibitem{meyer2013}
Bertrand Meyer.
\newblock Theory of programs.
\newblock In Bertrand Meyer and Martin Nordio, editors, {\em Proceedings of LASER summer schools 2007, 2008, 2013, 2014, Lectures on Software Engineering, Lecture Notes in Computer Science}, volume 8987. Springer, 2014.

\bibitem{milne}
Robert Milne and Christopher Strachey.
\newblock {\em A Theory of Programming Language Semantics)}.
\newblock Routledge, 2015.

\bibitem{morgan1990}
Carroll Morgan.
\newblock {\em Programming from specifications}.
\newblock Prentice-Hall, Inc., 1990.

\bibitem{morgan}
Carroll Morgan.
\newblock {\em Programming from Specifications, Second edition}.
\newblock Prentice Hall, 1998.

\bibitem{isabelle}
Tobias Nipkow, Markus Wenzel, and Lawrence~C. Paulson.
\newblock {\em Isabelle/HOL: a proof assistant for higher-order logic}.
\newblock Springer, 2002.

\bibitem{roscoecspnew}
A.W. Roscoe.
\newblock {\em Understanding Concurrent Systems: Communicating Sequential Processes}.
\newblock Springer, London, 2010.

\bibitem{weber2025}
Reto Weber.
\newblock {PRISM proofs repository (operational from 1 March 2025)}, February 2025.
\newblock Available at: \url{https://github.com/CI-CSE/PRISM}.

\bibitem{wirth1971}
Niklaus Wirth.
\newblock Program development by stepwise refinement.
\newblock {\em Communications of the ACM}, 14(4):221--227, 1971.

\bibitem{wirth1973}
Niklaus Wirth.
\newblock {\em Systematic Programming: An Introduction}.
\newblock Prentice Hall PTR, USA, 1973.

\end{thebibliography}

\end{document}